\documentclass[12pt]{iopart}
\usepackage{iopams} 
\input epsf
\expandafter\let\csname equation*\endcsname\relax
 \expandafter\let\csname endequation*\endcsname\relax
 \usepackage[fleqn]{amsmath}
\usepackage{amsopn}
\usepackage{epsfig}
\usepackage{subfig}
\usepackage{graphics,psfrag,rotating}
\usepackage{graphicx}
\usepackage{dcolumn}
\usepackage{bm}
\usepackage{epstopdf}
\usepackage{color}
\usepackage[usenames,dvipsnames,svgnames]{xcolor}
\usepackage[colorlinks=true,
      linkcolor=red,
      urlcolor=gray,
      citecolor=blue]{hyperref}
 \usepackage[]{cite}

\def\3nab{\tilde{\nabla}}

\def \t {\tilde}

\def\hsp5{\hspace{5mm}}

\def\case#1/#2{\textstyle\frac{#1}{#2}}

\def\ber {\begin{eqnarray}}
\def\eer {\end{eqnarray}}
\def\bea {\begin{eqnarray}}
\def\eea {\end{eqnarray}}

\def\bc {\begin{center}}
\def\ec {\end{center}}
\def\case#1/#2{\frac{#1}{#2}}

\newcommand{\bw}{\begin{widetext}}
\newcommand{\ew}{\end{widetext}}

\newcommand{\be}{\begin{equation}}
\newcommand{\bse}{\begin{subequation}}
\newcommand{\ese}{\end{subequation}}
\newcommand{\ee}{\end{equation}}
\newcommand{\eei}{\end{eqnarray}\indent\indent}
\newcommand{\ba}{\begin{array}}
\newcommand{\ea}{\end{array}}
\newcommand{\bal}{\begin{eqnarray}}
\newcommand{\eal}{\end{eqnarray}} 
\newcommand{\n}{{}^{(3)}\nabla}

\def\case#1/#2{\textstyle\frac{#1}{#2} }

\newcommand{\bver}{\begin{verbatim}}
\newcommand{\ever}{\end{verbatim}}

\begin{document}
\title{Covariant density and velocity perturbations of the quasi-Newtonian  cosmological model in $f(T)$ gravity}
\author{Heba Sami$^{1}$\footnote{
hebasami.abdulrahman@gmail.com},  Shambel Sahlu$^{2,3}$,  Amare Abebe$^{1}$ and  Peter K. S. Dunsby $^{4}$}
\address{$^{1}$ Center for Space Research, North-West University, South Africa}
\address{$^{2}$  Astronomy and Astrophysics  Research Development Department, Entoto Observatory and Research Center, Ethiopian Space Science and Technology Institute, Ethiopia}
\address{$^{3}$ Department of Physics, College of Natural and Computational Science, Wolkite University, Ethiopia}
\address{$^{4}$ Cosmology and Gravity Group, Department of Mathematics and Applied Mathematics,
University of Cape Town, Rondebosch 7701, Cape Town, South Africa.}
\date{\today}

\begin{abstract}

We investigate classes of shear-free cosmological dust models with irrotational fluid flows within the framework of $f(T)$ gravity. In particular, we use the $1 + 3$ covariant
formalism and present the covariant linearised evolution and constraint equations describing such models. We then derive the integrability conditions describing a consistent evolution of the linearised field equations of these {\it quasi-Newtonian} universes in the $f(T)$ gravitational theory. Finally, we derive the evolution equations for the density and velocity perturbations of the quasi-Newtonian universe. We explore the behaviour of the matter density contrast for two models - $f(T)= \mu T_{0}(T/T_{0})^{n}$ and the more generalised case, where $f(T)= T+ \mu T_{0} (T/T_{0})^{n}$, with and without the application of the quasi-static approximation.  Our numerical solutions show that these $f(T)$ theories can be suitable alternatives to study the background dynamics, whereas the growth of energy density fluctuations change dramatically from the expected $\Lambda$CDM behaviour even for small deviations away from the general relativistic limits of the underlying $f(T)$ theory. Moreover, applying the so-called quasi-static approximation yields exact-solution results that are orders of magnitude different from the numerically integrated solutions of the full system, suggesting that these approximations are not applicable here.
\end{abstract}
 
 \maketitle

\section{Introduction}
In 1998 \cite{riess1998observational, perlmutter1998discovery}, observational evidence for an accelerating universe was discovered within the framework of a Friedmann-Lema\^{i}tre-Robertson-Walker (FLRW) cosmology. The only way to explain this phenomenon is to introduce an additional dark component to the total energy density \cite{copeland2006dynamics}. However, the physical properties of Dark Energy (DE) is still not well understood. Moreover, there are many problems that need to be explained such as the inhomogeneity problem and how the primordial fluctuations seeded the formation of structure on large-scales. There are exist several approaches to the theoretical description of these problems.  One of these is a modified gravity theories, which provide the very natural gravitational alternative for dark energy and they are extremely attractive in the applications for late-time acceleration \cite{nojiri2007introduction}. One of these modified gravity theories is  $f(T)$ gravity \cite{liu2012energy, paliathanasis2018stability, capozziello2001geometric,sahlu2019accelerating, sahlu2019chaply,sahlu2020scalar}. In the simple case $f(T) = T$, where T is the torsion scalar. The $f(T)$ theory can be directly reduced to the teleparallel equivalent of general relativity (TEGR) \cite{tsamparlis1979cosmological}. The idea of the teleparallel gravity (TG) was  originally proposed in 1928 by Einstein after the formulation of general relativity (GR) to address the unification of gravity with  electromagnetism by introducing the notion of tetrad (vierbin) field together with the suggestion of absolute parallelism \cite{einstein1925unified,einstein1928riemannian,einstein1928new,unzicker2005translation}. In this theory, a set of tetrad fields $e_{a}(x)^{\mu}$ are considered to be the dynamical object instead of the metric $g_{ab}$. While (GR) uses the well-known torsionless Levi-Civita connection, TG uses Weitzenb\"{o}ckconnection \cite{turnbull1924invarianten,3} that has no curvature but instead torsion. From Einstein-Cartan-Kibble-Sciama theory of
gravitation \cite{sciama1962analogy}, torsion is an alternative to curvature to describe the gravitational interaction \cite{hehl1995metric, carvalho2004torsion}.
From this view, the TG description of the gravitational interaction is completely equivalent to that of GR in some respects \cite{aldrovandi2012teleparallel}. In (GR), curvature is used to geometrize the gravitational interaction, in other words, the gravitational force is replaced by geometry and particle trajectories are determined by geodesics, rather than the force equation \cite{darabi2015geodesic}. In (TG), there is no notion of geodesics and the torsion  yields a true gravitational force which quite similar to the Lorentz force of electrodynamics \cite{3, arcos2004torsion}. Although at the background and perturbation levels TG is completely equivalent to (GR). $f(T)$ gravity has various
cosmological solutions which are consistent with the observational data \cite{krvsvsak2017variational, linder2010einstein, iorio2012solar}, especially at the cosmological background. According to these features, $f(T)$ gravity is assumed as a viable theory both at cosmological and at astrophysical scales. One of the significant advantages of $f(T)$ gravity is that its field equations are always of second-order in contrast with $f(R)$ gravity, where the field equations are governed by fourth-order equations. In \cite{li2011f}, it was shown that $T$ is not a local Lorentz scalar but the lack of this local Lorentz symmetry appears to be of little importance in (TG) when the Lagrangian is just $T$. However, it is not the case for the $f(T)$ generalisation of the (TG). If $T$ is not a local Lorentz invariant then $f(T)$ cannot be either, and this is considered to be one of the disadvantages of the $f(T)$ gravity. 

In this paper, we consider the covariant form of the field equations of $f(T)$ gravity to study linear cosmological perturbations \cite{li2011large}. There are two approaches to study the cosmological perturbations namely, the metric based approach \cite{lifshitz1946gravitational, bardeen1980gauge, kodama1984cosmological,bertschinger2000cosmological,bruni1992cosmological,dunsby1991gauge, dunsby1992covariant,gidelew2013beyond,hwang1991large,mukhanov1992theory} and the $1+3$ covariant approach \cite{ hawking1966perturbations, ellis1989covariant}. In the $1+3$  covariant approach, the perturbations are formulated using variables that are covariantly defined in the real universe, and are exactly gauge-invariant by construction \cite{challinor2000microwave}. This approach has been used recently to study the cosmological perturbations for different contexts of modified gravity and $GR$ \cite{abebe2012covariant, dunsby1991gauge, ellis1989covariant}. 

This paper is organised as follows: in Sections \ref{sec1}, \ref{sec2} and \ref{sec3}  respectively, we review the $1+3$ covariant approach, kinematics quantities in the presence of torsion and we provide the covariant form of the field equations in $f(T)$ gravity which are required to study the cosmological perturbations. In Sections \ref{sec4} and \ref{sec5} respectively, we study the quasi-Newtonian models in the $f(T)$ gravity and we derive the integrability conditions that describes a consistent evolution of the linearised field equations of the quasi- Newtonian universes. In Section \ref{sec6}, we define the gradient variables that describe the cosmological perturbations and derive the linear evolution equations for matter and torsion perturbations. In Section \ref{sec7}, we analyse the growth of the matter density contrast by considering the power-law $f(T)$ theory where $f(T)= \mu T_{0}(T/T_{0})^{n}$ and the more generalised case, where $f(T)= T+ \mu T_{0} (T/T_{0})^{n}$ by solving the whole system of perturbation equations numerically. We introduce the so-called quasi-static approximation to admit the approximated solutions on small scales. Section \ref{sec8} is devoted for discussions and conclusions.
\section{The $1+3$ covariant approach in $f(T)$ gravity} \label{sec1}
In this approach, space-time is split into space and time, where $1+3$ refers to the number of dimensions involved in each slice to investigate the deviation from homogeneity and isotropy of the Universe. The $4$-velocity field vector of the observer $u^{a}$ associated with the worldlines is defined as 
\be u^{a}= \dfrac{dx^{a}}{d\tau}, \hspace*{1cm} u^{a}u_{a}=-1\;,\ee
where $x^{a}$ is the worldline in terms of local coordinates and $\tau$ is proper time measured along the worldlines. In this approach the metric $g_{ab}$ is decomposed into the projected tensor $h_{ab}$ as follows:
\be\label{1+3}
g_{ab}= h_{ab}- u_{a}u_{b}\;, \mbox{with}~h^{a}_{~c}h^{c}_{~b}= h^{a}_{~b}, ~~ h^{a}_{~a}=3, ~~~h_{ab}u^{b}=0\;.
\ee

We consider the covariant form of $f(T)$ gravity to clearly show the equivalence between teleparallel gravity and General Relativity. This form of field equation is very advisable to define the covariant variables in a gauge-invariant formalism for the study of the cosmological perturbations. Instead of using the torsionless Levi-Civita connection $\Gamma^{c}_{ab}$ in general relativity, we use the curvature-less  Weitzenb\"{o}ckconnection connection  $\tilde{\Gamma}^{c}_{ab}$ in TG . The torsion tensor has different symmetry properties from the curvature case and it can be expressed as \cite{cai2016f, darabi2015geodesic}
\begin{eqnarray}
T^{c}_{ab}= e^{c}_{\mu}(\partial_{a} e^{\mu}_{b}- \partial_{b}e^{\mu}_{a})\;.
\end{eqnarray}
 Each vector $e_{a}$ is described by its components $e^{\mu}_{a}= 0,1,2,3,$ in a coordinate basis.
The contorsion tensor is expressed as \cite{darabi2015geodesic}
\begin{equation}\label{k}
K^{c}_{ab}= \tilde{\Gamma}^{c}_{ab}- \Gamma^{c}_{ab}\;.
\end{equation}
The teleparallel Lagrangian density is described by the torsion scalar as follows \cite{darabi2015geodesic}
\begin{equation}
T= S^{ab}_{d}T ^{d}_{ab}\;,
\end{equation}
where the super-potential term is given as \cite{paliathanasis2018stability,cai2016f, darabi2015geodesic}
\begin{equation}
S^{ab}_{d}= K^{ba}_{d}+ \delta^{a}_{d}T^{\sigma b}_{\sigma}- \delta^{b}_{d}T^{\sigma a}_{\sigma}\;,
\end{equation} 
and the contortion tensor can be rewritten as 
\begin{equation}
K^{ba}_{d}= -\dfrac{1}{2}\big( T^{ab_{d}}- T^{ba}_{d}- T^{ab}_{d}\big)\;.
\end{equation}
The modified teleparallel  action for $f(T)$ is given by \cite{linder2010einstein, darabi2015geodesic}
 \begin{equation}\label{actionfTB}
S_{f(T)} = \frac{1}{2\kappa}\int{{\rm d}^4x \,e\,\left[f(T)+2\mathcal{L}_{m}\right]}\;,
 \end{equation}
 where $e$ is the determinant of the tetrad field $e^{\mu}_{a}$ i.e., $\left(e = \mbox{det}|e^{\mu}_{a}| = \sqrt{-g}\right) $, and the coupling constant $\kappa = 8\pi G/c^4$ \footnote{From here on-wards, the geometric units convention where $8\pi G=c =1$.}. Varying the action in Eq. \eqref{actionfTB} with respect to the vierbein vector field $e^{\mu}_{a}$, we obtain 
\begin{equation}
\dfrac{1}{e} \partial_{b}(eS^{ba}_{A})f'{T} - e^{\lambda}_{A}T^{d}_{b\lambda} S^{ab}_{d}f'(T)+ S^{ba}_{A}\partial_{b}(T)f''(T)+\dfrac{1}{4}e^{a}_{A}f(T)= k\Theta^{a}_{A}\;,
\end{equation}
where $f'$ and $f''$ denote the differentiation with respect to $T$ and $\Theta^{a}_{A}$ is the matter energy momentum tensor, and all aindices on the manifold run over $0, 1, 2, 3$. From the relation between the   Weitzenb\"{o}ckconnection and Levi-Civita connection in Eq. \eqref{k}, one can write the Riemann tensor associated with the Levi-Civita connection and contorsion tensor as \cite{darabi2015geodesic}
\be
R^{d}_{ab}= \partial_{a}\Gamma ^{d}_{cb}- \partial_{b}\Gamma^{d}_{ca}+ \Gamma^{d}_{fa}\Gamma^{f}_{cb}- \Gamma^{d}_{bf}\Gamma^{f}_{ca}= \nabla_{a}K^{d}_{cb} -\nabla_{b}K^{d}_{ca}+ K^{d}_{fa}K^{f}_{cb}- K^{d}_{fb}K^{f}_{ca}\;,
\ee
and the Ricci scalar is given as 
\begin{equation}
R= -T+2\nabla^{a}T^{b}_{ab}= -T+2\nabla^{a}T_{a}\;.
\end{equation}
The field equations can be written as 
\begin{equation}\label{Gab}
G_{ab}= - \dfrac{1}{2}g_{ab} T-\nabla^{c}S_{bca}- S^{dc}_{a}K_{cdb}\;,
\end{equation}
where $G_{ab}= R_{ab}- \dfrac{1}{2} g_{ab} R$ is the Einstein tensor and $g_{ab}$ is the metric tensor. From Eq. \eqref{Gab}, we consider a covariant version of the field equations of $f(T)$ gravity with a clear analogy to Einstein's field equations as \cite{liu2012energy,li2011large, darabi2015geodesic}
\begin{equation}
 f'G_{ab}+\frac{1}{2} g_{ab} [f-f'T]-f''S_{ab}^{~~~c}\nabla_c T = \kappa^2\mathcal{T}_{ab}\;, \label{field1}
\end{equation}
where $\mathcal{T}_{ab}$  denotes the usual energy-momentum tensor of the matter fluid expressed as $\mathcal{T} ^b_{a} = \frac{1}{e}\frac{\delta ( e L_m)}{\delta e^{b}_{a}}\;.$
\section{Kinematics quantities in the presence of torsion}\label{sec2}
 All the kinematic quantities which  describe all the kinematic features of the fluid flow can be obtained from irreducible parts of the decomposed $\nabla_{a}u_{b}$ as \cite{carloni2010conformal,gidelew2013beyond}
\begin{equation}\label{definitionofu}
\tilde{\nabla}_{a}u_{b}=\frac{1}{3}\tilde{\theta} h_{ab}+\tilde{\sigma}_{ab}+\tilde{\omega}_{ab}-u_{a}\tilde{\dot{u}}_{b}\; ,
\end{equation}
where tilde terms here are referring to the torsion contribution. The volume rate of expansion of the fluid in the presence of torsion is given as 
\begin{equation}\label{a}
\tilde{\theta}= \theta -2u^{b} T_{b}\;.
\end{equation}
The rate of distortion of the matter flow  is given as
\begin{equation}\label{b}
\tilde{\sigma}_{ab}=\sigma_{ab}+2h^{c}_{\hspace{.1 cm a}}h^{d}_{\hspace{.1 cm b}}K^{e}_{[cd]}u_{e}\; ,
\end{equation}
and the skew-symmetric vorticity tensor 
\begin{equation}\label{d}
\tilde{\omega}_{ab}=\omega_{ab}+ 2h^{c}_{\hspace{.1 cm a}}h^{d}_{\hspace {.1 cm b}}K^{e}_{[cd]}u_{e}\; ,
\end{equation}
describes the rotation of the fluid relative to a non-rotating frame.
The relativistic acceleration vector is given as
\begin{equation}\label{c}
\tilde{\dot{u}}_{a}=\dot{u}_{a}+u^{b}K^{e}_{ab}u_{e}\; .
\end{equation}
The general expression for the Raychaudhuri equation is given by 
\begin{equation}\label{raych}
\tilde{\dot{\theta}} = \tilde{\nabla}^{a} \tilde{\dot{u}}_{a}-\dfrac{1}{3}\theta^{2} - \tilde{\sigma}^{cb}\tilde{\sigma}_{cb}- \tilde{\omega}^{cb}\tilde{\omega}_{cb}- R_{cb}u^{c}u^{b}-2u^{b}T^{d}_{cb}\big( \dfrac{1}{3}h^{c}_{d}\tilde{\theta}-\tilde{\sigma}^{c}_{d}-\tilde{\omega}^{c}_{d}-u^{c}\tilde{\dot{u}}_{d}\big)\;.
\end{equation}
$\tilde{\omega}_{cb} = 0=\tilde{\sigma}_{cb}$ in the case of non-rotational and shear free fluids  and from the covariant approach of the field equation, $R_{cb}u^cu^b = 1/2\left(\rho +3p\right)$ for relativistic fluid  \cite{ellis1999cosmological, castaneda2016some}. Then the Raychaudhuri Eq.\eqref{raych} is rewritten as
\begin{eqnarray}\label{qq1}
 \dot{\tilde{{\theta}}} = \tilde{\nabla}^a\dot{\tilde{u}}_a -\frac{1}{3}{\theta}^2 - \frac{1}{2}\left(\rho + 3p\right) - \frac{2}{3
 }u^bT_{b}\tilde{{\theta}}\;.
\end{eqnarray}
The inner product of the torsion and four-velocity vectors of the fluid $u^bT_b$ is vanished identically For a space-like torsion vector \cite{pasmatsiou2017kinematics}. Therefore, Eq. \eqref{a} and Eq. \eqref{c} become $\tilde{{\theta}} = {\theta}$ and  $\dot{\tilde{u}}_a = \dot{u}_a$ respectively.
Then, from the result of Eq. \eqref{qq1}, we obtain
\begin{equation}\label{Raychaudhuri}
 \dot{{\theta}} = -\frac{1}{3} \theta^2-\frac{1}{2}\left(\rho+3p\right)+\tilde{\nabla}^a\dot{u}_a\;,
\end{equation}
and this equation is the same as the usual Raychaudhuri equation which is presented in 
\cite{dunsby1992covariant,li2011large, ellis1999cosmological,ehlers2007ak}. Now we consider the perturbations evolution in Friedmann universe,  the torsion scalar 
\begin{eqnarray}
T = -\frac{2{\theta}^2}{3} - \frac{4}{3}{\theta} \tilde{\nabla}^av_a\;.
\end{eqnarray}
The linearised thermodynamic quantities in the presence of torsion are the energy density $\rho_{T}$, the pressure $p_{T}$, the energy flux $q^{T}_{a}$ and the anisotropic pressure $\pi^{T}_{ab}$, respectively given by
\begin{eqnarray}\label{18}
&&\rho_{T}= -\dfrac{1}{f'}\Big[ (f'-1)\rho_{m}+\dfrac{1}{2} (f-f'T)\Big]\;,  \\&& \label{19}
p_{T}= -\frac{1}{f'}\Big[ (f'-1)p_{m}-\dfrac{1}{2} (f-f'T)\Big]+\dfrac{2f'' \dot{T}}{3f'}({\theta}+ 
\tilde{\nabla}^{a}v_a)\;,  \\
&&\label{20} 
q^{T}_{a}= -\frac{1}{f'}\Big[ (f'-1)q^{m}_{a}- \dfrac{2}{3} f''{\theta}\tilde{\nabla}_{a}T\Big]\;, \\&& \label{21}
\pi^{T}_{ab}= -\frac{1}{f'}\Big[(f'-1)\pi^{m}_{ab}- f''\dot{T}(\sigma_{ab}+\tilde{\nabla}_av_b)\Big]\;.
\end{eqnarray}
 The total effective energy density, isotropic pressure, anisotropic pressure and heat flux of standard matter and torsion combination are defined as
\begin{equation}\label{thermodynamicsquantities}
\rho \equiv\rho_{m}+ \rho_{T}, \hspace{.4cm} p \equiv p_{m}+ p_{T}, \hspace{.4cm} \pi_{ab}\equiv \pi^{m}_{ab}+ \pi^{T}_{ab}, \hspace{.4cm} q_{a}\equiv q^{m}_{a}+ q^{T}_{a}\;.
\end{equation}
From Eq. \eqref{field1}, the Freiedmann equations of the effective fluid are presented as follows:
\begin{eqnarray}\label{H1}
&&H^{2}= \dfrac{\rho_{m}}{3f'}-\dfrac{1}{6f'}(f-Tf')\;,   \hspace*{.4cm} 
2\dot{H} +3H^{2}= \dfrac{p_{m}}{f'}+\dfrac{1}{2f'}(f-Tf')+\dfrac{4f'' H \dot{T}}{f'}\;.\label{H2}
\end{eqnarray} 
One can directly obtain the corresponding thermodynamics quantities such as the effective energy density  and  the effective pressure of the fluid 
\begin{equation}\label{rho111}
\rho= \dfrac{\rho_{m}}{f'}-\dfrac{1}{2f'}(f-Tf')\;, \hspace*{.4cm}  p= \dfrac{p_{m}}{f'}-\dfrac{1}{2f'}(f-Tf')+\dfrac{2f'' H \dot{T}}{f’}\;,
\end{equation}
therefore, the Friedmann Eqs. \eqref{H1}  can be expressed as 
\begin{eqnarray}
&&1= \tilde{\Omega}_{m}+\mathcal{X}\;,    \hspace*{.4cm} \dfrac{\dot{H}}{H^2}= \left( -\dfrac{3}{2}+ \dfrac{3w}{2}\tilde{\Omega}_{m}-\dfrac{3}{2}\mathcal{X}+3\mathcal{Y}\right)\; .\label{FR}
\end{eqnarray}
For simplicity, we have introduced the following dimensionless variables as presented in \cite{sahlu2020scalar} are
\begin{eqnarray}\label{defi1}
 \mathcal{X}= \dfrac{Tf'- f}{6H^{2}f'}\;, \hspace*{.4cm}  \tilde{\Omega}_{m}= \dfrac{\rho_{m}}{3H^{2}f'}= \dfrac{\Omega_{m}}{f'}\;, \hspace*{.4cm}
\mathcal{Y}= \dfrac{2\dot{T} f''}{3H f'}\;.
\end{eqnarray}
Here $\tilde{\Omega}_{m}$ is the fractional energy density of effective matter like fluid, $\Omega_{m}$ is the normalized energy density parameter of standard matter fluid and $\mathcal{X}$ is the fractional energy density of torsion fluid.\\
\newline
In the FLRW spacetime universe, the  Raychaudhuri Eq. \eqref{Raychaudhuri} that governs the expansion history of the Universe can be written as follows due to the resulting non-trivial field equations  \cite{carloni2008evolution}:
\be\label{22}
\dot{\theta}= -\frac{1}{3} \theta^{2}-\dfrac{1}{2f^{'}}\Big[ \rho_{m}+f -Tf^{'}+ 2f^{''}\theta \dot{T} + 2f^{''}\dot{T}\tilde{\nabla}^{a}v_{a}\Big]+ \tilde{\nabla}^{a}A_{a}\;.
\ee
\section{Covariant equations}\label{sec3}
Given a choice of $4$-velocity field $u^{a}$, the Ehlers-Ellis approach \cite{ehlers1993contributions, ellis1989covariant} employs only fully covariant quantities and equations with transparent physical and geometric meaning \cite{maartens98}. In such a treatment for any scalar quantity $X$ in the background we have
$$\tilde{\nabla}_{a}X=0\; ,$$
thus, by virtue of the Stewart-Walker Lemma \cite{stewart1974perturbations}, any quantity is considered to be gauge-invariant if it vanishes in the background. Therefore, the FLRW background is characterised by the following dynamics, kinematics and gravito-electromanetics  \cite{abebe2016integrability}:
\be\label{1}
\tilde{\nabla}_{a}\rho=0=\tilde{\nabla}_{a}p=\tilde{\nabla}_{a}{\theta}\;, \quad A_{a}=0=\omega_{a}=q_{a}\;,\quad  \pi_{ab}=0=\sigma_{ab}=E_{ab}=H_{ab}\;,
\ee
where $E_{ab}$ and $H_{ab}$ are the gravito-electromagnetic fields responsible for tidal forces and gravitational waves. They are the ``gravito-electric'' and ``gravito-magnetic'' components of the Weyl tensor $C_{abcd}$  defined from the Riemann tensor $R^a_{bcd} $ as 
\begin{eqnarray}\label{weyl}
&&C^{ab}{}_{cd}=R^{ab}{}_{cd}-2g^{[a}{}_{[c}R^{b]}{}_{d]}+\frac{R}{3}g^{[a}{}_{[c}g^{b]}{}_{d]}\;,\\&& 
E_{ab}\equiv C_{agbh}u^{g}u^{h}\;,  \hspace*{.1cm} H_{ab}\equiv\dfrac{1}{2}\eta_{ae}{}^{gh}C_{ghbd}u^{e}u^{d}\;.
\end{eqnarray}
The covariant linearised evolution equations in the general case are given by \cite{maartens98, abebe2016integrability, maartens1997density} 
\begin{eqnarray}\label{4000}
&&\dot{{\theta}}= -\dfrac{1}{3}{\theta}^{2} - \dfrac{1}{2}(\rho+ 3p)+ \tilde{\nabla}_{a}A^{a}\;,  \hspace*{.2cm} \label{5}
\dot{\rho}_{m}= -\rho_{m}{\theta} -\tilde{\nabla}^{a}q^{m}_{a}\; ,\\
&&\label{6000}
\dot{q}^{m}_{a}= -\dfrac{4}{3}{\theta} q^{m}_{a}- \rho_{m} A_{a}\; ,  \hspace*{.2cm} 
\label{10}
\dot{\sigma}_{ab}=-\dfrac{2}{3}{\theta} \sigma_{ab}-E_{ab} + \dfrac{1}{2}\pi_{ab}+ \tilde{\nabla}_{\langle a}A_{b\rangle}\; ,\\
&&\label{9}
\dot{\omega}^{\langle a\rangle}= -\dfrac{2}{3}{\theta} \omega^{a}- \dfrac{1}{2}\eta^{abc}\tilde{\nabla}_{b}A_{c}\; ,\\
&&\label{7}
\dot{E}^{\langle ab\rangle}= \eta^{cd\langle a}\tilde{\nabla}_{c} H^{\rangle b}_{d}- {\theta} E^{ab} - \dfrac{1}{2}\dot{\pi}^{ab}-\dfrac{1}{2}\tilde{\nabla}^{\langle a}q^{b\rangle}- \dfrac{1}{6}{\theta} \pi^{ab}\; ,\\
&&\label{8}
\dot{H}^{\langle ab\rangle}= -{\theta} H^{ab}- \eta^{cd\langle a}\tilde{\nabla}_{c}E^{\rangle b}_{d}+ \dfrac{1}{2} \eta^{cd\langle a}\tilde{\nabla}_{c}\pi^{\rangle b}_{d}\;.
\end{eqnarray}
These evolution equations propagate consistent initial data on some initial $(t = t_{0})$
hypersurface $S_{0}$ uniquely along the reference time-like  consistency \cite{abebe2015irrotational}. They are constrained by the following linearised equations \cite{maartens98, abebe2016integrability, maartens1997density}:
\begin{eqnarray}\label{11}
&&C^{ab}_{0}\equiv E^{ab}- \tilde{\nabla}^{\langle a}A^{b \rangle}- \dfrac{1}{2} \pi^{ab}=0\; , \hspace*{.4cm} 
\label{12}
C^{a}_{1}\equiv \tilde{\nabla}_{b}\sigma^{ab}- \eta^{abc}\tilde{\nabla}_{b}\omega_{c}- \dfrac{2}{3}\tilde{\nabla}^{a}{\theta}+ q^{a}=0\; ,\\
&&\label{13}
C_{2}\equiv \tilde{\nabla}^{a}\omega_{a}=0\; , \hspace*{.4cm} 
\label{14}
C^{ab}_{3}\equiv\eta_{cd(}\tilde{\nabla}^{c}\sigma^{d}_{b)}+\tilde{\nabla}^{\langle a}\omega^{b\rangle}-H^{ab}=0\; ,\\
&&\label{15}
C^{a}_{5}\equiv\tilde{\nabla}_{b}E^{ab}+\dfrac{1}{2}\tilde{\nabla}_{b}\pi^{ab}-\dfrac{1}{3}\tilde{\nabla}^{a}\rho+\dfrac{1}{3}{\theta} q^{a}=0\; , \\&&
\label{16}
C^{a}_{b}\equiv\tilde{\nabla}_{b}H^{ab}+(\rho +p)\omega^{a}+\dfrac{1}{2}\eta^{abc}\tilde{\nabla}_{b}q_{a}=0\; .
\end{eqnarray}
These constraints restrict the initial data to be specified  and they must remain satisfied on any hypersurface $S_{t}$ for all comoving time $t$.
\section{Quasi-Newtonian spacetimes}\label{sec4}
There being no proper Newtonian limit for GR on cosmological scales, recent works on so-called quasi-Newtonian cosmologies \cite{maartens98,maartens1998newtonian,van1998quasi} have shown that gravitational physics can be studied to a good approximation \cite{gidelew2013beyond}. The importance of investigating the Newtonian limit for general relativity on cosmological contexts is that, there is a viewpoint that cosmology is essentially a Newtonian affair, with the relativistic theory only needed for examination of some observational relations. Most of the astrophysical calculations on the formation of large-scale structure in the universe rely on such a limit \cite{van1998quasi}. In \cite{van1998quasi}, a covariant approach to cold matter universes in quasi-Newton has been developed and it has been applied and extended in \cite{maartens1998newtonian} in order to derive and solve the equations governing density and velocity perturbations. This approach revealed the existence of integrability conditions in GR.\\
\newline
If a comoving  $4$-velocity $\hat{u}^{a}$ is chosen such that, in the linearised form 
\begin{equation}\label{29}
\hat{u}^{a}= u^{a}+\hat{v}^{a}, \hspace{.3cm} v_{a}u^{a}=0, \hspace{.3cm} v_{a}v^{a}<<1\;,
\end{equation}
the dynamical, kinematic and gravito-electromagnetic quantities Eq. \eqref{1} undergo transformation.\\
Here $v^{a}$ is the relative velocity  of the comoving frame with respect to the  observer in the quasi-Newtonian frame, defined such that it vanishes in the background. In other words, it is a non-relativistic peculiar velocity.
Quasi-Newtonian cosmological models are irrotational, shear-free dust spacetimes characterised by \cite{maartens98, abebe2016integrability}:
\begin{eqnarray}\label{24}
&& p_{m}=0\;, \hspace*{.3cm} q^{m}_{a}=\rho_{m} v_{a}\;, \hspace*{.3cm} \pi^{m}_{ab}=0\;,  \hspace{.3cm} 
\label{254}\omega_{a}=0\;, \hspace*{.3cm}  \sigma_{ab}=0\;.
\end{eqnarray}
Therefore, the evolution equations \eqref{4000} - \eqref{8} for this class of spacetimes can be written as 
\begin{eqnarray}
&&\dot{{\theta}}= -\dfrac{1}{3}{\theta}^{2} - \dfrac{1}{2}(\rho+ 3p)+ \tilde{\nabla}_{a}A^{a}\;, \hspace*{.2cm} 
\label{5a}
\dot{\rho}_{m}= -\rho_{m}{\theta} -\tilde{\nabla}^{a}q^{m}_{a}\; ,\\
&&\label{6000a}
\dot{q}^{m}_{a}= -\dfrac{4}{3}{\theta} q^{m}_{a}- \rho_{m} A_{a}\; , \hspace*{.2cm} 
\label{7a}
\dot{E}^{\langle ab\rangle}= \eta^{cd\langle a}\tilde{\nabla}_{c} H^{\rangle b}_{d}- {\theta} E^{ab} - \dfrac{1}{2}\dot{\pi}^{ab}-\dfrac{1}{2}\tilde{\nabla}^{\langle a}q^{b\rangle}- \dfrac{1}{6}{\theta} \pi^{ab}\; ,\\
&&\label{8a}
\dot{H}^{\langle ab\rangle}= -{\theta} H^{ab}- \eta^{cd\langle a}\tilde{\nabla}_{c}E^{\rangle b}_{d}+ \dfrac{1}{2} \eta^{cd\langle a}\tilde{\nabla}_{c}\pi^{\rangle b}_{d}\; .
\end{eqnarray}
Due to the vanishing of the shear in the quasi-Newtonian frame, Eq. \eqref{10} is turned into a new constraint 
\begin{equation}\label{45}
E_{ab} =  \dfrac{1}{2}\pi^{\phi}_{ab} + \tilde{\nabla}_{\langle a}A_{b\rangle}\; ,
\end{equation}
and using the identity in Eq. \eqref{a1} for any scalar $\varphi$, Eq. \eqref{9} can be simplified as
\begin{equation}\label{46}
\eta^{abc} \tilde{\nabla}_{a}A_{c}=0 \Rightarrow A_{a}= \tilde{\nabla}_{a}\varphi\;,
\end{equation}
where  $\varphi$ is the covariant relativistic generalisation of the Newtonian potential.
The shear-free $\sigma_{ab}=0$ and irrotational condition $\omega=0$ and the gravito electromagnetic(GEM) constraint Eq. $\eqref{14}$ result in the silent constraint $H_{ab}=0$\;.
Thus there is no gravitational radiation, which further justifies the term $‘quasi-Newtonian’$. And  Eq.$\eqref{16}$ show that $q_{a}$ is irrotational and thus $v_{a}$
\begin{equation}\label{26}
\dfrac{1}{2}\eta^{abc}\tilde{\nabla}_{b}q_{a}=0=\dfrac{1}{2}\eta^{abc} \rho_{m}\tilde{\nabla}_{b} q_{a}\;,
\end{equation}
it follows that for a vanishing vorticity, there exists a velocity potential $\psi$ such that
$v_{a}= \tilde{\nabla}_{a}\psi\;.$

\section{Integrability conditions}\label{sec5}
A constraint equation $C_{A}=0$  evolves consistently with the evolution equations in the sense that \cite{maartens1997linearization,maartens1998newtonian,maartens98,abebe2016integrability}
\begin{equation}\label{34}
\dot{C}_{A}= F^{A}_{B}C^{B}+G^{A}_{Ba}D^{a}C^{B}\; ,
\end{equation}
where $F$ and $G$ depend on the kinematic, dynamical and gravito-electromagnetic quantities but not their derivatives \cite{maartens98,abebe2016integrability,Sami:2018lxt}. It has been shown that the non-linear models are generally inconsistent if the silent constraint  $H_{ab}=0$ is imposed, but that the linear models are consistent. Thus, so a simple approach to the integrability conditions for quasi-Newtonian cosmologies follows from showing that these models are in fact a sub-class of the linearized silent models.\\
This can  happen by using the transformation between the quasi-Newtonian and comoving frames.\\
\newline
The transformed linearised kinematics, dynamics and gravito-electromagnetic quantities from the quasi-Newtonian frame to the comoving frame are given as follows \cite{maartens98,abebe2016integrability,van97, Sami:2018lxt}:
\begin{eqnarray}\label{35}
&&\tilde{\Theta} =\theta +\tilde{\nabla}^{a}v_{a}\; ,\hspace*{.2cm} 
\label{36}
\tilde{A}_{a}= A_{a}+ \dot{v}_{a}+\dfrac{1}{3}\theta v_{a}\; ,\hspace*{.2cm} 
\label{37}
\tilde{\omega}_{a}= \omega_{a}- \dfrac{1}{2}\eta_{abc}\tilde{\nabla}^{b}v^{c}\; ,\\&&
\label{38}
\tilde{\sigma}_{ab}= \sigma_{ab}+ \tilde{\nabla}_{\langle a}\hat{v}_{b\rangle}\; , \hspace*{.2cm} \label{39}
\tilde{\rho}= \rho, \hspace{.2cm} \tilde{p}=p, \hspace{.2cm} \tilde{\pi}_{ab}= \pi_{ab}, \hspace{.2cm} \tilde{q}^{T}_{a}= q^{T}_{a}\; ,\\&& 
\label{40}
\tilde{q}^{m}_{a}= q^{m}_{a}-(\rho_{m} +p_{m})v_{a}\; ,\hspace*{.2cm} 
\label{41}
\tilde{E}_{ab}= E_{ab}, \hspace{.2cm} \tilde{H}_{ab}= H_{ab}\; .
\end{eqnarray}
\subsection*{The first and second integrability conditions}
From Eq. \eqref{45},  we need to ensure its consistent propagation at all epochs and in all spatial hypersurfaces. The differentiating this equation with respect to a cosmic time $t$ and  and together with Eqs. \eqref{21}, \eqref{12} and \eqref{7a} the identity  of \eqref{a2}, one  we obtain
\begin{eqnarray}\label{447}
&&-\bigg(\frac{f'''\dot{T}^2}{f'} +\frac{f''\ddot{T}}{f'} - \frac{f''^2\dot{T}^2}{f'^2} + \frac{{\theta}}{3} \frac{f''\dot{T}}{f'}\bigg)\tilde{\nabla}_av_b - \tilde{\nabla}_{a}\tilde{\nabla}_{b} \Big(\dot{\varphi}+\dfrac{1}{3}{\theta}\Big) \\ \nonumber && - \Big(\dot{\varphi} +\dfrac{1}{3}{\theta} +\frac{f''\dot{T}}{f'} \Big)\tilde{\nabla}_{a}\tilde{\nabla}_{b}\varphi=0\;.
\end{eqnarray}
This equation is the first integrability condition for quasi- Newtonian cosmologies in $f(T)$ gravity and it is a generalisation of the one obtained in \cite{maartens1998newtonian}. Eq. \eqref{447} reduces to an identity for the generalized van Elst-Ellis condition  \cite{van1998quasi,maartens98,abebe2016integrability, Sami:2018lxt} as
\begin{equation}\label{661}
 \dot{\varphi} +\dfrac{1}{3}{\theta} = - \frac{f''\dot{T}}{f'}\;.
\end{equation}
From Eq. \eqref{5a} with the time evolution of the modified van Elst-Ellis condition Eq. \eqref{661}, we obtain the covariant modified Poisson equation in $f(T)$ gravity as follows as
\begin{eqnarray}\label{710}
&& \tilde{\nabla}^{2}\varphi + 3\ddot{\varphi} + {\theta}\dot{\varphi} =  \dfrac{1}{2f'}\rho_{m}+\dfrac{1}{2f'}\Big(f -Tf'+ 2f''\theta \dot{T}  \\ \nonumber &&+ 2f''\dot{T}\tilde{\nabla}^{a}v_{a}-6f''\dot{T}^{2}-6f''\ddot{T}+\frac{f''^2\dot{T}^2}{f'}+2f''\dot{T}\theta\Big)\;.
\end{eqnarray}
 For the case of $f= T$, Eq. \eqref{710} reduces to the one obtained in \cite{maartens98}. By taking the gradient of Eq. \eqref{661}, one gets
\begin{equation}\label{73}
\tilde{\nabla}_{a}\dot{\varphi} + = - \dfrac{1}{3}\tilde{\nabla}_{a} {\theta}- \tilde{\nabla}_{a}\Big(\frac{f''\dot{T}}{f'}\Big)\;,
\end{equation}
using the identity \eqref{a3}, the above Eq. \eqref{73} can be written as
\begin{equation}\label{77}
\Big(\tilde{\nabla}_{a}\varphi\Big)^{\cdot}=-\dfrac{1}{3}\tilde{\nabla}_{a} {\theta}-\tilde{\nabla}_{a} \Big(\frac{f''\dot{T}}{f'}\Big) - \dfrac{1}{3}{\theta}\tilde{\nabla}_{a}\varphi+\dot{\varphi}A_{a}\;,
\end{equation}
using Eq. \eqref{46} together with  the shear-free constraint Eq. \eqref{12}, one can obtain the evolution equation of the $4$-acceleration $A_{a}$ as
\begin{equation}\label{89}
\dot{A_{a}} = -\dfrac{\rho_{m}v_{a}}{2f'}-\bigg(\frac{f''{\theta}}{3f’}+\frac{f'''\dot{T}}{f'}-\frac{f''^2\dot{T}}
{f'^2}\bigg)\tilde{\nabla}_a{T} - \frac{f''}{f'}\tilde{\nabla}_a{\dot{T}} - \Big(\dfrac{1}{3}{\theta}-\dot{\varphi}\Big)A_{a}\;,
\end{equation}
 and it reduces to the one obtained in \cite{maartens98}.
To check for the consistency of the constraint Eq. \eqref{45} on any spatial hyper-surface of constant time $t$, we take the divergence of Eq. \eqref{45} ad by using the identity \eqref{a4}, we get
\begin{equation}\label{96}
\tilde{\nabla}^{b}E_{ab}= \dfrac{1}{2}\tilde{\nabla}^{b}\pi_{ab}+\dfrac{1}{2}\tilde{\nabla}^{2}(\tilde{\nabla}_{a}\varphi)+\dfrac{1}{6}\tilde{\nabla}_{a}(\tilde{\nabla}^{2}\varphi)+ \dfrac{1}{3}(\rho- \dfrac{1}{3} {\theta}^{2})\tilde{\nabla}_{a}\varphi\;,
\end{equation}
by using the constraint Eq.\eqref{15} with the identity \eqref{a10} it follows that:
\begin{equation}\label{100}
-\tilde{\nabla}^{b}\pi_{ab}+\dfrac{1}{3}\tilde{\nabla}_{a}\rho-\dfrac{1}{3}{\theta} q_{a}=\dfrac{2}{3}\tilde{\nabla}_{a}(\tilde{\nabla}^{2}\varphi)+ \dfrac{2}{3}(\rho- \dfrac{1}{3} {\theta}^{2})\tilde{\nabla}_{a}\varphi\;.
\end{equation}
By using Eqs. \eqref{21}, \eqref{thermodynamicsquantities} and \eqref{12}, one obtains
\begin{eqnarray}\label{secondin}
 &&\hspace{-0.5 cm} {\tilde{\nabla}_{a}\rho^{m}}-{2}\left({\rho_m} + \frac{1}{2}(f-Tf')- \dfrac{f'}{3} {\theta}^{2}\right)\tilde{\nabla}_{a}\varphi - \dfrac{2f'}{3}{\theta}\tilde{\nabla}_{a}{\theta} - {2f'}\tilde{\nabla}_{a}(\tilde{\nabla}^{2}\varphi)   \\ \nonumber &&= 3{f''\dot{T}}\tilde{\nabla}^{b}(\tilde{\nabla}_{a}v_b)  +\Big(\dfrac{\rho^{m} f''}{f'}- \dfrac{f''f}{2f'}\Big)\tilde{\nabla}_{a}T  \;,
\end{eqnarray}
which is the second integrability condition, The left and right hand sides of Eq. \eqref{secondin} refer to GR and non-GR contributions. For GR, the right hand side vanishes when $f= T$ and this matches the result obtained in \cite{maartens1997density} for the second integrability condition. By taking the gradient of Eq. \eqref{661} and using Eq. \eqref{12}, one can obtain the peculiar velocity:
\begin{equation}\label{118}
v_{a} = -\dfrac{2}{\rho^{m}}\bigg[ \tilde{\nabla}_{a}\dot{\varphi}+\bigg(\frac{f'''\dot{T}}{f'}-\frac{f''^2\dot{T}}{f'^2}\bigg)\tilde{\nabla}_a{T} + \frac{f''}{f'}\tilde{\nabla}_a{\dot{T}}\bigg]\;,
\end{equation}
which evolves according to
\begin{equation}\label{124}
\dot{v}_{a}+\dfrac{1}{3}{\theta} v_{a}= -A_{a}\;.
\end{equation}
\section{Cosmological perturbations} \label{sec6}
In Section \ref{sec5}, we showed how imposing special restrictions to the linearised perturbations of FLRW universes in the quasi-Newtonian setting result in the integrability conditions. These integrability conditions imply velocity and acceleration propagation equations resulting from the generalised van Elst-Ellis condition for the acceleration potential in $f(T)$ gravity. In this section, we  obtain the velocity and density perturbations via these propagation equations, thus generalizing GR results obtained in \cite{maartens98}.
\subsection{Definition of vector gradient variables}
The covariant vector gradient variable $D^{m}_{a}$ for the total matter fluid and the volume expansion of the fluid can be defined respectively as follows
\begin{eqnarray}\label{Da}
&&D^{m}_{a}= \dfrac{a\tilde{\nabla}_{a}\rho_{m}}{\rho_{m}}\;,  \hspace{.4cm} Z_{a}= a\tilde{\nabla}_{a}\theta\; .\label{Za}
\end{eqnarray}
Those two gradient variables are the key to examine the evolution equation for the matter density perturbations.

Now we define an extra key variables resulting from the spatial gradient of gauge-invariant quantities which are connected with torsion fluid for $f(T)$ gravity, we define  $\mathcal{F}_{a}$ for the torsion density fluid and $\mathcal{B}_{a}$ for the torsion momentum density respectively as follow.  
\begin{eqnarray}\label{Fa}
&&\mathcal{F}_{a}= a\tilde{\nabla}_{a}T\;, \hspace{.4cm}
 \mathcal{B}_{a}= a\tilde{\nabla}_{a}\dot{T}\;.\label{Ba}
\end{eqnarray}
We also define the comoving acceleration $\mathcal{A}_{a}$ and the covariant vector gradient variable $V^{m}_{a}$ for the velocity inhomogeneity of the matter as follows
\begin{eqnarray}\label{{Aa}}
&&\mathcal{A}_{a}=aA_{a}\; , \hspace{.4cm}
\label{Va}
V^{m}_{a}= av_{a}\; .
\end{eqnarray}
\subsection{Linear evolution equations}
In the following we derive the linear evolution for each covariant variables. The system of equations governing the evolution of the variables defined in the previous subsection are given as follows

\begin{eqnarray}
 && \dot{Z}_{a} =  -\left(\frac{2\theta}{3}  +\frac{f''\dot{T}}{f'} \right)Z_a -\frac{1}{2f'}\rho_m D^m_a + \bigg(\frac{f''}{2f'^2}\rho_m + \frac{f'' f}{2f'^2} -\frac{f'''\dot{T}\theta}{f'}  +\frac{f''^2\dot{T}\theta}{f'^2}\bigg)\mathcal{F}_a \\ \nonumber && - \frac{f''\theta}{f'}\mathcal{B}_a + \tilde{\nabla}^{2}\mathcal{A}_{a}+ \bigg(-\frac{1}{3} \tilde{{\theta}^{2}}-\frac{1}{2f'}\Big( \rho_m + (f-Tf’) +2f''\dot{T}\theta \Big)\bigg) \mathcal{A}- \frac{f''\dot{T}}{f'}\tilde{\nabla}^2V^m_a\;, \label{z0}
\\
&&\dot{D}^{m}_{a} = -Z_a - {\theta} \mathcal{A}_a- \tilde{\nabla}^{2}V^{m}_{a} \;,\label{matterdenityfluid}\\
 && \dot{\mathcal{A}}_{a}= \dot{\varphi}\mathcal{A}_{a} - \dfrac{\rho_{m}}{2f'}V^m_a-\bigg(\frac{f''{\theta}}{3f'} +\frac{f'''\dot{T}}{f'}-\frac{f''^2\dot{T}}
{f'^2}\bigg)\mathcal{F}_a -\frac{f''}{f'}\mathcal{B}_a \label{accelaration} \;, \\&&
   \dot{\mathcal{F}_{a}}-\mathcal{B}_{a}- \dot{T} \mathcal{A}_{a}=0\;, \label{densitytorsion}\\&&
   \dot{\mathcal{B}_{a}}-\dfrac{\dddot{T}}{\dot{T}} \mathcal{F}_{a}- \ddot{T} \mathcal{A}_{a}=0 \;, \label{momuntumdensity}\\&&
\dot{V}^{m}_{a}+\mathcal{A}_{a}=0\;.
 \end{eqnarray}
\subsection{Definition of scalar gradient variables}
The vector gradient variables we have defined so far contain both a scalar and a vector parts.
However, the formation of structures on large scales follows a spherical clustering mechanism, and this could only be studied through the scalar parts. Therefore, we extract the scalar parts of the perturbation vectorial gradients of these quantities by applying a local decomposition \cite{ellis1990density}
\begin{eqnarray}\label{localdecom}
a\tilde{\nabla}{a}X= X_{ab}=\dfrac{1}{3}h_{ab}X+ \Sigma^{X}_{ab}+X_{[ab]}\; ,
\end{eqnarray}
where $\Sigma^{X}_{ab} = X_ {(ab)}-3 h_{ ab} X$ describes shear whereas $X _{[ab]}$ describes the vorticity. When extracting the scalar contribution the vorticity term vanishes.\\
Based on the above decomposition, by applying the comoving differential operator $a\tilde{\nabla}_{a}$ to Eqs. \eqref{Da}-\eqref{Va}, our scalar variables can be given to linear order as
\begin{eqnarray}
&&\Delta_{m}= a\tilde{\nabla}^{a}D^{m}_{a}= \dfrac{a^{2} \tilde{\nabla}^{2}\rho_{m}}{\rho_{m}}\; \;,\quad Z= a\tilde{\nabla}^{a}Z_{a}= a^{2}\tilde{\nabla}^{2}{\theta}\;,\quad V^{m}= a\tilde{\nabla}V^{m}_{a}= a^{2}\tilde{\nabla}^{a}v_{a}\;,\nonumber\\
&&\mathcal{F}= a\tilde{\nabla}^{a}\mathcal{F}_{a}= a^{2}\tilde{\nabla}^{2}T\;,\quad \mathcal{A}= a\tilde{\nabla}^{a}\mathcal{A}_{a}= a^{2}\tilde{\nabla}^{a}A_{a}\;,\quad \mathcal{B}= a\tilde{\nabla}^{a}\mathcal{B}_{a}= a^{2}\tilde{\nabla}^{2}\dot{T}\; \nonumber\\
\end{eqnarray}
\subsection{first- and second- order evolution equations}
Due to the above  definitions of the scalar gradient variables, here we present first- and second-order evolution equations to demonstrate the growth of perturbations with cosmological red shift as
\begin{eqnarray}\label{00}
 && \dot{Z}  +\left(\frac{2\theta}{3}  +\frac{f''\dot{T}}{f'} \right)Z +\frac{1}{2f'}\rho_m \Delta^m - \bigg(\frac{f''}{2f'^2}\rho_m + \frac{f'' f}{2f'^2} -\frac{f'''\dot{T}\theta}{f'}   +\frac{f''^2\dot{T}\theta}{f'^2}\bigg)\mathcal{F} \nonumber\\&&+ \frac{f''\theta}{f'}\mathcal{B} +\frac{f''\dot{T}}{f'}\tilde{\nabla}^2V_m- \tilde{\nabla}^{2}\mathcal{A}\nonumber\\&&- \bigg(-\frac{1}{3} \tilde{{\theta}^{2}}-\frac{1}{2f'}\Big( \rho_m + (f-Tf’) +2f''\dot{T}\theta \Big) \bigg) \mathcal{A}=0\;, 
 \\
&&  \dot{\Delta}^{m} +Z + {\theta} \mathcal{A}+ \tilde{\nabla}^{2}V_{m}=0\label{matterdenityfluid}\;, \\
 &&\dot{\mathcal{A}}+(\dfrac{1}{3}\theta+ \dfrac{f'' \dot{T}}{f'})\mathcal{A} + \dfrac{\rho_{m}}{2f'}V_m+\bigg(\frac{f''{\theta}}{3f'} +\frac{f'''\dot{T}}{f'}-\frac{f''^2\dot{T}}{f'^2}\bigg)\mathcal{F} +\frac{f''}{f'}\mathcal{B}=0 \label{accelaration}\;, \\&&
\dot{\mathcal{F}}- \mathcal{B} - \dot{T} \mathcal{A}=0\label{densitytorsion}\;, \\&&
\dot{\mathcal{B}}- \dfrac{\dddot{T}}{\dot{T}} \mathcal{F}- \ddot{T} \mathcal{A}=0\label{momuntumdensity}\;,\\&&
\dot{V}^{m}+\mathcal{A}=0 \label{ddot10100}\;, \\&&
 \ddot{\Delta}_{m} +\left(\frac{2\theta}{3}  + \frac{f''\dot{T}}{f'} \right)\dot{\Delta}_m -\frac{1}{2f'}\rho_m \Delta_m  +\Bigg(\frac{1}{3} \tilde{{\theta}^{2}} +\dfrac{1}{f'}(\rho_{m} +f-Tf')\Bigg)\dot{V}_m  \nonumber \\&&- \dfrac{{\theta}\rho_{m}}{2f’}V_m- \dfrac{2f'' \theta}{f'}\dot{\mathcal{F}}  +\Bigg(\frac{f''}{2f'^2}\rho_m + \frac{f''f}{2f'^2}  -\frac{f''{\theta^2}}{3f'} -\dfrac{2f''' \dot{T} \theta}{f'}+ \dfrac{2f''^{2} \dot{T}\theta}{f'^2}\Bigg)\mathcal{F}=0 \;, \\&& \label{ddotf}
\ddot{\mathcal{F}} +\frac{f'' \dot{T}}{f'}\dot{\mathcal{F}} +\bigg(\frac{f'' \dot{T}{\theta}}{3f'}+\frac{f'''\dot{T}^2}{f'}- \frac{f''^2\dot{T}^2}{f'^2}-\dfrac{\dddot{T}}{\dot{T}}\bigg)\mathcal{F} \nonumber \\&&+ \left( 2\ddot{T} - \dfrac{\dot{T}\theta}{3}\right)\dot{V}_m  + \dfrac{\rho_{m} \dot{T}}{2f’}V_m=0\;,\\&& \label{ddotv}
\ddot{V}_{m} + \frac{1}{3}\theta\dot{V}_m- \dfrac{\rho_{m}}{2f'}V_m -\frac{f''}{f'}\dot{\mathcal{F}}  -\bigg(\frac{f''{\theta}}{3f'} +\frac{f'''\dot{T}}{f'}-\frac{f''^2\dot{T}} {f'^2}\bigg)\mathcal{F}=0 \;.
\end{eqnarray}
\subsection{Harmonic decomposition}
The above evolution Eq. \eqref{00} - \eqref{ddotv} can be thought of as a coupled system of harmonic
oscillator differential equations of the form \cite{gidelew2013beyond,carloni2006gauge}
\begin{eqnarray}
\ddot{X}+ A\dot{x}+ BX= C(Y,\dot{Y})\; , \label{Hdecomposition}
\end{eqnarray}
where $A$, $B$ and $C$ are independent of $X$ and they represent friction (damping), restoring and source forcing terms respectively.
To solve Eq. \eqref{Hdecomposition}, a separation of variables is applied such that
$$X(x,t)= X(\vec{x}) X(t), \hspace*{1cm} Y(x,t)= Y(\vec{x})Y(t)\; .$$
\newline
Since the evolution equations obtained so far are complicated to be solved, the harmonic decomposition approach is applied to these equations using the eigenfunctions and the corresponding wave number for these equations, therefore we write
$$X= \sum_{k}X^{k}Q_{k}(\vec{x})\; , \hspace*{1cm} Y=\sum_{k}Y^{k}(t)Q_{k}(\vec{x})\;,$$
where $Q_{k}(x)$ are the eigenfunctions of the covariantly defined spatial Laplace-Beltrami
operator \cite{gidelew2013beyond,carloni2006gauge}, such that
$$\tilde{\nabla}^{2}Q= -\dfrac{k^{2}}{a^{2}}Q\; .$$
The order of the harmonic (wave number) is given by $k=\dfrac{2\pi a}{\lambda}\; ,$ where $\lambda$ is the physical wavelength of the mode. The eigenfunctions $Q$ are covariantly constant, ie  $\dot{Q}_{k}(\vec{x})=0\; .$ Therefore, the first-order evolution Eqs. \eqref{00} - \eqref{ddotv}  become 
\begin{eqnarray}\label{000}
 && \dot{Z}_k +\left(\frac{2\theta}{3}  +\frac{f''\dot{T}}{f'} \right)Z_k + \frac{1}{2f'}\rho_m \Delta^m_k + \dfrac{f'' \theta}{f’} \mathcal{\dot{F}}_k- \bigg(\frac{f''}{2f'^2}\rho_m + \frac{f'' f}{2f'^2} -\frac{f'''\dot{T}\theta}{f'}  +\frac{f''^2\dot{T}\theta}{f'^2}\bigg)\mathcal{F}_k \nonumber \\ && + \bigg(-\frac{1}{3} \tilde{{\theta}^{2}}-\frac{1}{2f'}\Big( \rho_m + (f-Tf’) \Big)-\dfrac{k^2}{a^2} \bigg) \dot{V}^m_k-  \dfrac{f'' \dot{T} k^2}{a^2 f'} V^m_k=0\;, 
 \\&&  \dot{\Delta}^{m}_k +Z_k - {\theta} \dot{V}^m_k - \dfrac{k^2}{a^2}V^{m}_k=0\label{matterdenityfluid0}\;, \\
 &&\dot{\mathcal{A}}_{k}+(\dfrac{1}{3}\theta+ \dfrac{f'' \dot{T}}{f'})\mathcal{A}_{k} + \dfrac{\rho_{m}}{2f'}V^m_{k}+\bigg(\frac{f''{\theta}}{3f'} +\frac{f'''\dot{T}}{f'}-\frac{f''^2\dot{T}}{f'^2}\bigg)\mathcal{F}_{k} +\frac{f''}{f'}\mathcal{B}_{k}=0 \label{accelaration}\;, \\&&
\dot{\mathcal{F}}_{k}- \mathcal{B} _{k}- \dot{T} \mathcal{A}_{k}=0\label{densitytorsion}\;, \\&&
\dot{\mathcal{B}}_{k}- \dfrac{\dddot{T}}{\dot{T}} \mathcal{F}_{k}- \ddot{T} \mathcal{A}_{k}=0\label{momuntumdensity}\;,\\&&
\dot{V}^{m}_{k}+\mathcal{A}_{k}=0 \label{ddot10100}\;, \\&&
 \ddot{\Delta}^{m}_{k} +\left(\frac{2\theta}{3}  + \frac{f''\dot{T}}{f'} \right)\dot{\Delta}^m_{k} -\frac{1}{2f'}\rho_m \Delta^m_{k}  +\Bigg(\frac{1}{3} \tilde{{\theta}^{2}}+\dfrac{1}{f'}(\rho_{m} +f-Tf')\Bigg)\dot{V}^m_{k} \nonumber\\&&- \dfrac{{\theta}\rho_{m}}{2f’}V^m_{k}- \dfrac{2f'' \theta}{f'}\dot{\mathcal{F}}_{k} +\Bigg(\frac{f''}{2f'^2}\rho_m + \frac{f''f}{2f'^2}  -\frac{f''{\theta^2}}{3f'} -\dfrac{2f''' \dot{T} \theta}{f'}+ \dfrac{2f''^{2} \dot{T}\theta}{f'^2}\Bigg)\mathcal{F}_{k}=0 \;, \\&& \label{ddotf}
\ddot{\mathcal{F}}_{k} +\frac{f'' \dot{T}}{f'}\dot{\mathcal{F}}_{k} +\bigg(\frac{f'' \dot{T}{\theta}}{3f'}+\frac{f'''\dot{T}^2}{f'}- \frac{f''^2\dot{T}^2}{f'^2}-\dfrac{\dddot{T}}{\dot{T}}\bigg)\mathcal{F}_{k} \nonumber\\&&+ \left( 2\ddot{T} - \dfrac{\dot{T}\theta}{3}\right)\dot{V}^m_{k} + \dfrac{\rho_{m} \dot{T}}{2f’}V^m_{k}=0\;,\\&& \label{ddotv}
\ddot{V}^{m}_{k} + \frac{1}{3}\theta\dot{V}^m_{k}- \dfrac{\rho_{m}}{2f'}V^m_{k} -\frac{f''}{f'}\dot{\mathcal{F}}_{k}  -\bigg(\frac{f''{\theta}}{3f'} +\frac{f'''\dot{T}}{f'}-\frac{f''^2\dot{T}} {f'^2}\bigg)\mathcal{F}_{k}=0 \;.
\end{eqnarray}
Then, we will study the growth  of the matter density contrast with cosmological redshift. To do this we applied the transformation technique  to make the redshift dependent instead of cosmic time. 
Therefore,  our evolution equations  can be written as follow:
\begin{eqnarray}\label{0000}
 && Z' - \dfrac{1}{H(1+z)}\left(\frac{2\theta}{3}  +\frac{f''\dot{T}}{f'} \right)Z -\dfrac{1}{2f'H(1+z)} \rho_m \Delta_m + \dfrac{f'' \theta}{f’} \mathcal{F}' \nonumber\\&&+ \dfrac{1}{H(1+z)}\bigg(\frac{f''}{2f'^2}\rho_m + \frac{f'' f}{2f'^2} -\frac{f'''\dot{T}\theta}{f'}  +\frac{f''^2\dot{T}\theta}{f'^2}\bigg)\mathcal{F} \nonumber \\ && + \bigg(-\frac{1}{3} \tilde{{\theta}^{2}}-\frac{1}{2f'}\Big( \rho_m + (f-Tf’) \Big)-\dfrac{k^2}{a^2} \bigg) V'_m+ \dfrac{f'' \dot{T} k^2}{ H a^2 (1+z) } V_m=0\;, 
 \\&&  \Delta'_m -\frac{1}{H(1+z)}Z - {\theta} V'_m+ \dfrac{k^2}{a^2 	H (1+z)}V_{m}=0\label{matterdenityfluid0}\;, \\
 &&\mathcal{A}'-\dfrac{1}{3}\theta V'_{m}- \dfrac{\rho_{m}}{2f' H (1+z)}V_m+ \dfrac{f''}{f'} \mathcal{F}' \nonumber\\&&- \dfrac{1}{H(1+z)}\bigg(\frac{f''{\theta}}{3f'} +\frac{f'''\dot{T}}{f'}-\frac{f''^2\dot{T}}
{f'^2}\bigg)\mathcal{F}=0 \label{accelaration0}\;, \\
&&\mathcal{F}'+ \frac{1}{H(1+z)}\mathcal{B}+ \frac{\dot{T}} {H(1+z)}\mathcal{A}=0\label{densitytorsion0}\;, \\&&
\mathcal{B}'+ \dfrac{\dddot{T}}{\dot{T}H (1+z)} \mathcal{F} + \frac{\ddot{T}}{H(1+z)} \mathcal{A}=0\label{momuntumdensity0}\;,\\&&
V'_{m}- \frac{1}{H(1+z)}\mathcal{A}=0 \label{ddot1010000}\;, \label{ddotd0}
\end{eqnarray}
\begin{eqnarray}
&& \Delta''_{m} -\frac{1}{(1+z)}\left(\frac{1}{2}+ \frac{f''\dot{T}}{f'H}  \right) \Delta'_m- \frac{\rho_m}{2H^2 f'(1+z)^2} \Delta_m  \nonumber\\&&-\frac{1}{H(1+z)}\Bigg(\frac{1}{3} \tilde{{\theta}^{2}}+\dfrac{1}{f'}(\rho_{m} +f-Tf') \Bigg)V’_m  -\frac{1}{H^2 (1+z)^2}\bigg(\dfrac{ \theta \rho_{m}}{2f'} \bigg)V_m  \nonumber\\&&+\frac{1}{H(1+z)} \Big(\dfrac{2f'' \theta}{f'}\Big) \mathcal{F}'+\frac{1}{H^2 (1+z)^2} \Bigg(\frac{f''}{2f'^2}\rho_m + \frac{f''f}{2f'^2}  -\frac{f''{\theta^2}}{3f'} \nonumber\\&&-\dfrac{2f''' \dot{T} \theta}{f'} + \dfrac{2f''^{2} \dot{T}\theta}{f'^{2}}\Bigg)\mathcal{F} =0\;,\\&& \label{ddotf0}
\mathcal{F}''+ \frac{1}{(1+z)} \left( \frac{3}{2}- \frac{f'' \dot{T}}{f' H}\right)\mathcal{F}' +\frac{1}{H^2 (1+z)^2}\bigg(\frac{f'' \dot{T}{\theta}}{3f'}+\frac{f'''\dot{T}^2}{f'}- \frac{f''^2\dot{T}^2}{f'^2}- \dfrac{\dddot{T}}{\dot{T}}\bigg)\mathcal{F}\nonumber \\&& 
- \frac{1}{H(1+z)}\left( 2\ddot{T} -\dfrac{\theta \dot{T}}{3}\right)V'_m +\dfrac{\rho_{m} \dot{T}}{2f' H^2(1+z)^2}V_m =0\;,\\&& \label{ddotv0}
V''_{m} +\frac{1}{2(1+z)} V'_m- \dfrac{\rho_{m}}{2f' H^2(1+z)^2}V_m +\frac{f''}{f' H(1+z)}\mathcal{F}' \nonumber\\&&- \frac{1}{H^2 (1+z)^2}\bigg(\frac{f''{\theta}}{3f'} +\frac{f'''\dot{T}}{f'}-\frac{f''^2\dot{T}} {f'^2}\bigg)\mathcal{F} =0\;.
\end{eqnarray} 
For more simplicity, we introduce here some quantities such as:
\begin{eqnarray}\label{b1}
&&\phi = \frac{f''}{f'} \Bigg(\frac{\rho_{m}}{2f'} + \frac{f}{2f'}  -\dfrac{f''' \dot{T} \theta}{f''}+  \dfrac{f'' \dot{T}\theta}{f'}\Bigg)\;, \quad \beta = \frac{f''}{f'} \Bigg(\frac{\rho_{m}}{2f'} + \frac{f}{2f'}  -\frac{\theta^2}{3} -\dfrac{2f''' \dot{T} \theta}{f''}+ \dfrac{2f'' \dot{T}\theta}{f'}\Bigg) \;, \nonumber\\
&& \zeta =\bigg(\frac{f'' \dot{T}{\theta}}{3f'}+\frac{f'''\dot{T}^2}{f'}- \frac{f''^2\dot{T}^2}{f'^2}- \dfrac{\dddot{T}}{\dot{T}}\bigg) \;, \quad
 \eta = \bigg(\frac{f''{\theta}}{3f'} +\frac{f'''\dot{T}}{f'}-\frac{f''^2\dot{T}} {f'^2}\bigg)\;.
 \end{eqnarray}
By using the Friedmann Eq. \eqref{H2} and the introduced dimensionless variables from Eq. \eqref{defi1}, then we rewrite the more generalised form of the evolution equations as follows: 
\begin{eqnarray}\label{000001}
 && Z' - \dfrac{1}{(1+z)}\left(2 +\frac{3\mathcal{Y}}{2} \right)Z-\dfrac{3 H\tilde{\Omega}_{m}}{2(1+z)} \Delta_m + \dfrac{f'' \theta}{f’} \mathcal{F}'+ \dfrac{\phi}{H(1+z)}\mathcal{F} \nonumber \\ && +3H^2 \bigg(-1 -\frac{\tilde{\Omega}_{m}}{2}+\mathcal{X}- \dfrac{k^2}{3 H^2 a^2} \bigg) V'_m+ \dfrac{3\mathcal{Y}k^2}{ 2a^2 (1+z) } V_m=0\;, 
 \\&&  \Delta'_m -\frac{1}{H(1+z)}Z - {\theta} V'_m+ \dfrac{k^2}{a^2 	H (1+z)}V_{m}=0\label{matterdenityfluid0}\;, \\
 &&\mathcal{A}'-H V'_{m} - \dfrac{3 H\tilde{\Omega}_{m}}{2 (1+z)}V_m+ \dfrac{f''}{f'} \mathcal{F}'- \dfrac{\eta }{H(1+z)}\mathcal{F}=0 \label{accelaration0}\;, \\&&
\mathcal{F}'+ \frac{1}{H(1+z)}\mathcal{B}+ \frac{\dot{T}} {H(1+z)}\mathcal{A}=0\label{densitytorsion0}\;, \\&&
\mathcal{B}'+ \dfrac{\dddot{T}}{\dot{T}H (1+z)} \mathcal{F}+ \frac{\ddot{T}}{H(1+z)} \mathcal{A}=0\label{momuntumdensity0}\;,\\&&
V'_{m}- \frac{1}{H(1+z)}\mathcal{A}=0 \label{ddot1010000}\;, \\&&
 {\Delta}''_m  - \frac{1}{2(1+z)}\left(1  + 3\mathcal{Y}\right){\Delta}'_m- \frac{3\bar{\Omega}_m }{2(1+z)^2}\Delta_m  -\frac{3H}{(1+z)}\Bigg(1+\bar{\Omega}_m -2\mathcal{X})\Bigg){V}'_m \nonumber\\&&- \frac{9\bar{\Omega}_mH}{2(1+z)^2}V_m + \frac{9H\mathcal{Y}}{\dot{T}(1+z)}\mathcal{F}'+ \frac{\beta}{H^2 (1+z)^2} \mathcal{F}= 0 \;, \label{non-quasi1} \\&& 
 \mathcal{F}''- \frac{3}{2(1+z)}(\mathcal{Y}-1)\mathcal{F}'+\frac{\zeta}{H^2 (1+z)^2}\mathcal{F}  -\frac{1}{H(1+z)}\left( 2\ddot{T} - \dfrac{\dot{T}\theta}{3}\right)V'_m  \nonumber\\&&+ \dfrac{3\bar{\Omega}_m\dot{T}}{2(1+z)^2}  V_m = 0\;,\\&& \label{non-quasi2}
V''_m  +\frac{1}{2(1+z)}{V}'_m- \dfrac{3\bar{\Omega}_m }{2(1+z)^2}V_m +\frac{3\mathcal{Y}}{2(1+z)\dot{T}}\mathcal{F}'- \frac{\eta}{H^2 (1+z)^2}\mathcal{F}  = 0\;.\label{non-quasi3301}
\end{eqnarray}
 For further analysis, in this part we are going to apply the quasi-static approximation to our evolution equations \eqref{non-quasi1} - \eqref{non-quasi3301}. In this approximation, we assume very slow temporal fluctuations in the perturbations of both the torsion energy density and its momentum compared with the fluctuations of the matter energy density.   Therefore, terms involving time derivatives for torsion fluid are neglected, i.e., $\mathcal{F}^{'}=\mathcal{F}^{''}\approx 0$.
Then , the second-order evolution equations \eqref{non-quasi1} - \eqref{non-quasi3301} for quasi-static approximations yield as
\begin{eqnarray}\label{quasi1}
&& \mathcal{F}-\dfrac{H(1+z)}{\zeta}(2\ddot{T} + \dot{T} H) V'_{m} + \dfrac{3 \bar{\Omega}_{m} \dot{T} H^2}{2\zeta} V_m = 0\;, \\ &&  \label{quasi2}
V''_m +\frac{1}{2(1+z)}\bigg(1-\frac{2\eta}{H\zeta}\left( 2\ddot{T} - {\dot{T}H}\right)\bigg)V'_m- \dfrac{3\bar{\Omega}_m}{2(1+z)^2}\bigg(1-\frac{\eta \dot{T}}{\zeta} \bigg)V_m  = 0\;, \\ &&\label{quasi0}
 {\Delta}''_m - \frac{1}{2(1+z)}\left(1 + 3\mathcal{Y}\right){\Delta}'_m - \frac{3\bar{\Omega}_m }{2(1+z)^2}\Delta_m  -\frac{3H}{(1+z)}\Big((1+\bar{\Omega}_m -2\mathcal{X}) \nonumber\\&&- \frac{\beta}{3H^2 \zeta}\left( 2\ddot{T} - {\dot{T}H}\right) \Big){V}’_m  - \frac{3\bar{\Omega}_mH}{2(1+z)^2} \bigg(3+\frac{\beta \dot{T}}{H\zeta} \bigg)V_m =0\;. 
 \end{eqnarray}
In the  above evolution equations \eqref{000001} - \eqref{non-quasi3301} , we have to note that  the $k$ dependence appears in the first- order evolution equations and  disappeasr in  the second-order equations and this exactly the same  as the work presented  in  \cite{maartens1998covariant}  for GR. The GR  can be recovered for the case of $f(T) = T$, and we have 

\begin{eqnarray}
&& Z' - \dfrac{2}{(1+z)}Z-\dfrac{3 H\tilde{\Omega}_{m}}{2(1+z)} \Delta_m  +3H^2 \bigg(-1 -\frac{\tilde{\Omega}_{m}}{2}- \dfrac{k^2}{3 H^2 a^2} \bigg) V'_m=0\;, 
 \\&&   \Delta'_m -\frac{1}{H(1+z)}Z - {\theta} V'_m+ \dfrac{k^2}{a^2 	H (1+z)}V_{m}=0\label{matterdenityfluid0GR}\;, \\
 &&\mathcal{A}'-H V'_{m} - \dfrac{3 H\tilde{\Omega}_{m}}{2 (1+z)}V_m=0  \label{accelaration0GR}\;, \\&&
V'_{m}- \frac{1}{H(1+z)}\mathcal{A}=0 \label{ddot1010000GR}\;, \\&& 
  {\Delta}''_m  - \frac{1}{2(1+z)}{\Delta}'_m- \frac{3\bar{\Omega}_m }{2(1+z)^2}\Delta_m  -\frac{3H}{(1+z)}\Big(1+\bar{\Omega}_m \Big){V}'_m - \frac{9\bar{\Omega}_mH}{2(1+z)^2}V_m= 0 \label{GRdensity1}\;,\\&&
 V''_m    + \frac{1}{2(1+z)}{V}'_m  -  \dfrac{3{\Omega}_m }{2(1+z)^2} V_m= 0\;.\label{GRV}
\end{eqnarray}
In the following section, we explore  the solutions of the density and velocity contrast in GR and  $f(T)$ gravity models.

\section{Solutions}\label{sec7}
In this section we will solve the whole system of  perturbations equations we obtained so far  Eqs. \eqref{000001} - \eqref{non-quasi3301} to explore the growth of the matter density contrast in the GR context and for $f(T)$ gravity approach for non-quasi-static approximations and quasi-static approximation from Eqs. \eqref{quasi1} - \eqref{quasi0}. The exact solutions of the matter density contrast can be found in the quasi static approximations and the numerical solution will be presented in the non-quasi-static approximations as well. To find those solutions, we consider the power-law $f(T)$ model where $f(T)= \mu T_{0}(T/T_{0})^{n}$ and the more generalized model where $f(T)= T+ \mu T_{0} (T/T_{0})^{n}$. These models produce the accelerating expansion of the Universe without invoking the cosmological constant. The cosmological and spherical solutions in $f(T)$ gravity lead to various viable models that support cosmological observations along with the solar system tests. In this section we investigate the matter density contrast and the velocity  contrasts for non-quasi static and quasi-static approximations. We also show how these $f(T)$ gravity respond to the linear cosmological perturbations and formation of large-scale. We defined the normalized energy density for matter fluid as presented in \cite{sahlu2020scalar} \begin{equation}
\delta(z)=\frac{\Delta _m(z)}{\Delta (z_{in})} \;,                                                                                                                     
\end{equation}
where $\Delta _{in}$ is the initial value of $ \Delta_{m}(z)$ at $z_{in} = 1100$, since the variation of CMB temperature detected observationally in the order of $10^{-5}$ \cite{smoot1992structure} at $z\approx 1100$. In the same manner, we define normalized velocity contrast as
\begin{equation}
\nu(z)=\frac{V_m (z)}{V(z_{in})} \;,                                                                                                                     
\end{equation}
\subsection{The growth of matter and velocity-density fluctuations in GR limits}
Here, we analyse the growth of matter energy density contrasts $\delta(z)$ and the velocity contrast $\nu(z)$ with cosmic-time. 
We notice that, the second-order evolution equation \eqref{GRdensity1} is an opened system. While, the evolution equation for velocity Eq. \eqref{GRV} is a closed system, and easy to construct the exact solution of the velocity contrast. Then the exact solutions of the velocity contrast  yields  as
\begin{equation}
 V_m (z) = c_1\left(1+z\right)^{-1}+ c_2\left(1+z\right)^{\frac{3}{2}}\;.\label{solutionVGR}
\end{equation}
The integration constant $c_1$ and $c_2$  can be determined by the imposing initial conditions for plotting.
Those constants are worthy to find the exact solution for the density contrasts, and  we present  in the following as
\begin{eqnarray}
&& {\it c_1}=\frac{-2}{5} (1+z_{in}) \Big((1+z_{in}) \dot{V}_m(z_{in}) -\dfrac{3}{2} V_m(z_{in})\Big)\;, \nonumber\\&&{\it c_2}= \dfrac{2}{5 \sqrt{(1+z_{in}) }} \Big(\dot{V}_m(z_{in}) + \dfrac{V_m(z_{in})}{(1+z_{in})}\Big) \;.
\end{eqnarray}
Consequently, the second-order evolution equation of the matter density equation \eqref{GRdensity1} becomes a closed system by substituting the solutions of the velocity contrast Eq. \eqref{solutionVGR} and it's first order derivative.  Then, the exact solution is given as
\begin{eqnarray}
 && \Delta_m = \dfrac{1}{2} c_{1} (1+z)^{1/2}+3 c_{2} (1+z)^{3}+c_{3} (1+z)^{(3/4- \sqrt{33}/4)} + c_{4} (1+z)^{(3/4+ \sqrt{33}/4)}
  \;,\label{solutionGRd}
\end{eqnarray}
where $c_3$ and $c_4$ are the integration constants and they are given as
\begin{eqnarray}
&&{ \it c_{3, 4}}=\frac{\mp \frac{2}{\sqrt{33}}}{ (1+z_{in})^{\big(\frac{-1}{4} -\frac{\sqrt{33}}{4}\big)} } \Big\lbrace \dot{\Delta}_m(z_{in}) - \dfrac{\big(\frac{3}{4} \pm \frac{\sqrt{33}}{4}\big)}{(1+z_{in})} \Delta_m(z_in) \nonumber\\&&- \dfrac{ {\it c_{2}}\big(\frac{5}{4} \mp\frac{\sqrt{33}}{4}\big)}{2\sqrt{(1+z_{in})}} - 3{\it c_{1}} \Big(\frac{9}{4} \mp \frac{\sqrt{33}}{4}\Big) (1+z_{in})^{2}\Big\rbrace \;.
\end{eqnarray}
In the following figures we present  growth the matter density and velocity fluctuations with cosmological redshift in the GR approach. We set the initial conditions at $ V _{in}= V(z_{in}\simeq 1100)= 10^{-5}$ and $\dot{V}_{in}=\dot{V} (z_{in}=1100)= 0$ and $ \Delta _{in}= \Delta _{m}(z_{in}\simeq 1100)= 10^{-5}$ and $\dot{\Delta}_{in}=\dot{\Delta}_{m} (z_{in}=1100)= 0$. For the case when $f(T)=T$, $\delta(z)= \delta_{GR}(z)$ which coincides with TEGR and the results are exactly the same as GR. 
 \begin{figure}[h!]
 \begin{minipage}{0.45\textwidth}
\includegraphics[width=0.9\textwidth]{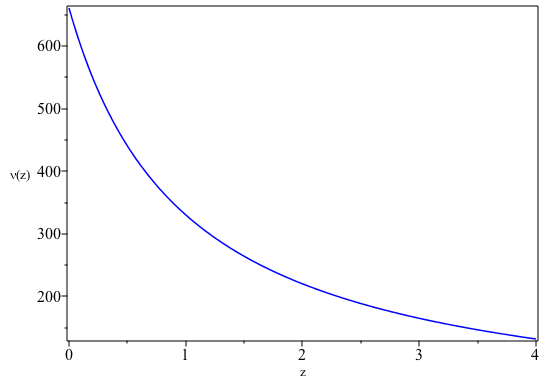}
    \caption{The growth of the velocity contrast  for Eq. \eqref{solutionVGR} (GR limits).}
    \label{fig:GRV}
\end{minipage} 
\qquad
 \begin{minipage}{0.45\textwidth}
\includegraphics[width=0.9\textwidth]{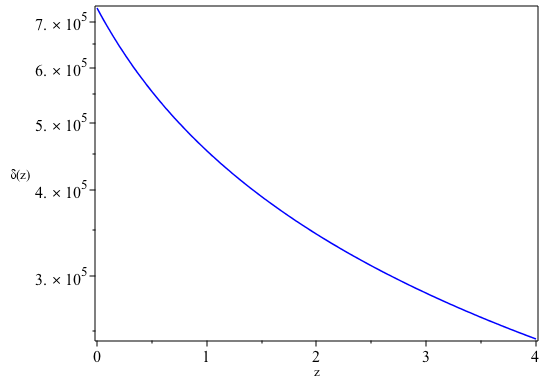}
   \caption{The growth of the density contrast for Eq. \eqref{solutionGRd} (GR limits).}
    \label{fig:GRd}
\end{minipage} 
 \end{figure}
 From this plots, we depict clearly the contribution of dust component of the universe for the fluctuations of matter density and velocity  are growing  with decreasing red shift. In the following section we will consider two paradigmatic $f(T)$ gravity models to clearly see the contributions of both (dust and torsion) fluids for the growth  of the fluctuations as well.   
\subsection{The growth of matter and velocity-density contrasts in $f(T)$ gravity models}
For more simplicity, we first find the following parameters in $f(T)$ gravity  models as 
\begin{eqnarray}&&
\phi =-\dfrac{3\mathcal{Y}}{8 \psi} \left( \dfrac{\bar{\Omega}_{m}}{2} -1- \mathcal{X} -2(n-2)\psi +\frac{3\mathcal{Y}}{2}\right)\;, \;\\&&
\beta =-\dfrac{\mathcal{Y}}{8 \psi} \left( \dfrac{3 \bar{\Omega}_{m}}{2} -3 \mathcal{X} -6 -12(n-2)\psi +9\mathcal{Y}\right) \;, \label{beta}
\\&&\zeta =  H^2 \left( \dfrac{3}{2} \mathcal{Y} (1-\dfrac{3}{2} \mathcal{Y}+2(n-2)\psi) -12 \psi^{2}  \right)\;, \label{zeta}
\\
 &&\eta = -\dfrac{\mathcal{Y}}{8H\psi} \left( 1-\frac{3}{2}\mathcal{Y}+2(n-2)\psi\right)\;,\label{eta}
  \\&& \dot{T} = -12H\dot{H} =-12H^3\psi\;, \\&&
\psi =\left( -\dfrac{3}{2}+ \dfrac{3w}{2}\tilde{\Omega}_{m}-\dfrac{3}{2}\mathcal{X}+3\mathcal{Y}\right)\;.
\end{eqnarray}
Consequently, the evolution  Eqs. \eqref{000001}- \eqref{non-quasi3301} for $f(T)$ gravity model are given as
\begin{eqnarray}\label{00000}
&& Z' - \dfrac{1}{(1+z)}\left(2 +\frac{3\mathcal{Y}}{2} \right)Z -\dfrac{3 H\tilde{\Omega}_{m}}{2(1+z)} \Delta_m - \dfrac{3 \mathcal{Y} }{8\psi H} \mathcal{F}'  \nonumber \\ &&-  \dfrac{3\mathcal{Y}}{8H \psi (1+z)}  \left( \dfrac{\bar{\Omega}_{m}}{2} -1- \mathcal{X} -2(n-2)\psi +\frac{3\mathcal{Y}}{2}\right)\mathcal{F} \nonumber \\ && +3H^2 \bigg(-1 -\frac{\tilde{\Omega}_{m}}{2}+\mathcal{X}- \dfrac{k^2}{3 H^2 a^2} \bigg) V'_m+ \dfrac{3\mathcal{Y}k^2}{ 2a^2 (1+z) } V_m=0\;, 
 \\&&  \Delta'_m -\frac{1}{H(1+z)}Z -3H V'_m+ \dfrac{k^2}{a^2 	H (1+z)}V_{m}=0\label{matterdenityfluid0}\;, \\
 &&\mathcal{A}'-H V'_{m} - \dfrac{3 H\tilde{\Omega}_{m}}{2 (1+z)}V_m- \dfrac{\mathcal{Y}}{8H^2 \psi} \mathcal{F}' \nonumber\\&&+ \dfrac{\mathcal{Y}}{8H^2\psi(1+z)} \left( 1-\frac{3}{2}\mathcal{Y}+2(n-2)\psi\right)\mathcal{F}=0 \label{accelaration0}\;, \\&&
\mathcal{F}'+ \frac{1}{H(1+z)}\mathcal{B}- \frac{12 H^2 \psi} {(1+z)}\mathcal{A}=0\label{densitytorsion0}\;, \\&&
\mathcal{B}'+ \dfrac{12H \psi^2}{ (1+z)} \mathcal{F}- \frac{36 H^3 \psi^2}{(1+z)} \mathcal{A}=0\label{momuntumdensity00}\;,\\&&
V'_{m}- \frac{1}{H(1+z)}\mathcal{A}=0\label{ddot10100000}\;, \\&&
 {\Delta}''_m  - \frac{1}{2(1+z)}\left(1  + 3\mathcal{Y}\right){\Delta}'_m - \frac{3\bar{\Omega}_m }{2(1+z)^2}\Delta_m  -\frac{3H}{(1+z)}\Bigg(1+\bar{\Omega}_m -2\mathcal{X})\Bigg){V}’_m  \nonumber\\&&- \frac{9\bar{\Omega}_mH}{2(1+z)^2}V_m -\frac{3\mathcal{Y}}{4H^2 \psi(1+z)}\mathcal{F}' \nonumber\\&&- \frac{\mathcal{Y}}{8H^2 \psi (1+z)^2}\left( \dfrac{3 \bar{\Omega}_{m}}{2} -3 \mathcal{X} -6 -12(n-2)\psi +9\mathcal{Y}\right)  \mathcal{F} = 0 \;, \label{non-quasi111} \\&& 
 \mathcal{F}''- \frac{3}{2(1+z)}(\mathcal{Y}-1)\mathcal{F}'+\frac{1}{ (1+z)^2} \left( \dfrac{3}{2} \mathcal{Y} (1-\dfrac{3}{2} \mathcal{Y}+2(n-2) \psi) -12 \psi^{2} \right)\mathcal{F}\\ \nonumber &&  -\frac{12 H^3 \psi }{(1+z)}\left( 1-6\psi\right){V}'_m  - \dfrac{18\bar{\Omega}_mH^3 \psi }{(1+z)^2}  V_m = 0\;,\\&& \label{non-quasi222}
V''_m  +\frac{1}{2(1+z)}V'_m- \dfrac{3\bar{\Omega}_m }{2(1+z)^2}V_m -\frac{\mathcal{Y}}{8 \psi H^3 (1+z)}\mathcal{F}'  \nonumber \\ &&+ \frac{\mathcal{Y}}{8H^3 (1+z)^2}  \left( 1-\frac{3}{2}\mathcal{Y}+2(n-2)\psi\right)\mathcal{F}  = 0\;.\label{non-quasi3333}
\end{eqnarray}

\subsection*{$f(T)$ power-law model}
In this subsection, we consider the paradigmatic power law model which is considered to be the simplest and compatible with the cosmic acceleration for $n > 1.5$ \cite{wei2012noether}, it is given as
\begin{eqnarray}\label{ftmodel}
f(T)= \mu T_{0}\left(\frac{T}{T_{0}}\right)^n\;,
\end{eqnarray}
where $\mu$and $n$ are dimensionless constants, and in background cosmology we have $T=-6H^2$  and $T_{0}= -6H_{0}^{2}$ is the present value of the torsion scalar. For $n=1$ this model reduces to the GR limit. We assume the scale factor $a(t)$ of the form
\begin{eqnarray}\label{scale}
 a= a_{0} \Big(\frac{t}{t_0}\Big)^m\;,
\end{eqnarray}
\newline 
where $m= \frac{2}{3(1+w)}$ is a positive constant with $w=0$ and normalized coefficient $a_{0}$ and $t_{0}$. The Hubble parameter is $H(z)= H_{0}h(z)$, where 
$ h(z)= \dfrac{2}{3} (1+z)^{3/2}\;.$
The background quantities $\mathcal{X}$, $\tilde{\Omega}_{m}$ and $\mathcal{Y}$ as defined in Eq. \eqref{defi1} become 
\begin{eqnarray} &&
 \mathcal{X}= \frac{1-n}{n}\;,  \hspace*{.4cm} 
\label{omega11}
\tilde{\Omega}_{m}= \frac{2n-1}{n}\;, ~\mbox{($n\ge 0.5$)}\;,\quad
\mathcal{Y}=  \frac{2(n-1)\Big(w(2n-1)-1\Big)}{n(5-4n)}\;.
\end{eqnarray}
To make sure that our evolution equations are dimensionless, we redefine the following normalized quantities as:
\begin{equation}\label{def11}
 Z =  H_{0}\mathcal{V} \;, \hspace*{.4cm}
\mathcal{F}= H_{0}^{2}F\;,  \hspace*{.4cm} 
V_{m}= \dfrac{1}{H_{0}} v\;,  \hspace*{.4cm} \mathcal{B}= H^3_{0} B\;.
\end{equation}
\subsection*{Case I: Solving the whole system for the power- law $f(T)$ model }
In this sub-subsection, we present the numerical results of the velocity and matter density fluctuations for power-law $f(T)$ gravity.  We start by solving the whole system of the first-order evolution equations \eqref{00000} - \eqref{ddot10100000}. We have evaluated the numerical solutions simultaneously to analyze the density fluctuations.  We set the initial conditions at $ V _{in}= V(z_{in}\simeq 1100)= 0$,  $  \mathcal{F} _{in}= \mathcal{F}(z_{in}\simeq 1100)= 0$ , $\mathcal{B} _{in}= \mathcal{B}(z_{in}\simeq 1100)= 0$, $\mathcal{A} _{in}= \mathcal{A}(z_{in}\simeq 1100)= 0$,  $Z _{in}= Z(z_{in}\simeq 1100)= 0$  and $\Delta _{in}= \Delta _{m}(z_{in}\simeq 1100)= 10^{-5}$. The numerical results are presented in the following Figs. \ref{Figuref2} - \ref{Figuref3}. For the case of $n = 1$, the numerical results of GR are recovered.   We have noticed the growth of the density contrast  for values of $n> 1.5 $, see Fig. \ref{Figuref3} and what we noticed here is that  only at this specific choice of the initial conditions, the $k$ dependence does not make any difference in the behaviour or the amplitude  of the density contrast for  long- and short wavelength (different values of $k$) . 
\begin{figure}[h!]
  \begin{minipage}{0.5\textwidth}
\includegraphics[width=0.9\textwidth]{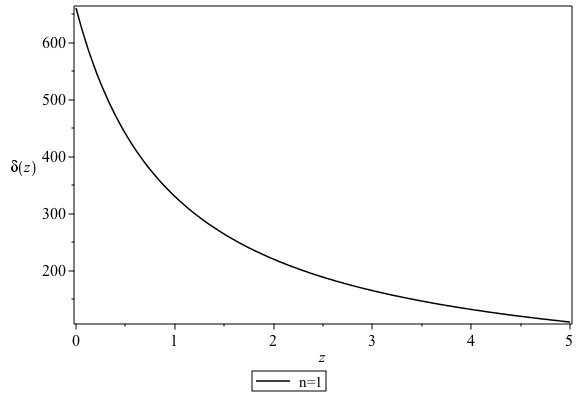}
   \caption{The growth of the density contrast versus cosmological redshift for the power law $f(T)$ gravity model for the first order equations  \eqref{00000} - \eqref{ddot10100000}  for $n = 1$.}
    \label{Figuref2}
\end{minipage} 
\qquad
\begin{minipage}{0.48\textwidth}
\includegraphics[width=0.9\textwidth]{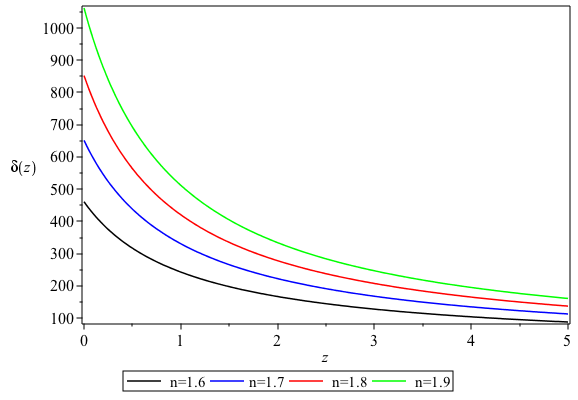}
        \label{fig:lol11}
    \caption{The growth of the density contrast versus cosmological redshift for the power law $f(T)$ gravity model for the first order equations  \eqref{00000} - \eqref{ddot10100000}  for $n>1.5$ and $k= 10^5$ and $k=0$.}
   \label{Figuref3}
\end{minipage}
\end{figure}
In the following, we will solve the whole system of the second-order perturbation equations,  we have evaluated the numerical solutions from Eqs. \eqref{non-quasi111} - \eqref{non-quasi3333}  simultaneously to analyze the density and velocity fluctuations with redshift. For further analysis, we study the growth of the matter density fluctuations with redshift and present the numerical results for different initial conditions as presented in \cite{abebe2013large}, to see how sensitive the results are to change the initial conditions:\\
\begin{itemize}
\item[I] $ V(z_{in}\simeq 1100)= \Delta _{m}(z_{in}\simeq 1100)= 10^{-5}$ and $\dot{V} (z_{in}=1100)= \dot{\Delta} _{m}(z_{in}\simeq 1100)= 0$,
\item[II] $ V(z_{in}\simeq 1100)= \Delta _{m}(z_{in}\simeq 1100)= 10^{-5}$ and $\dot{V} (z_{in}=1100)= \dot{\Delta} _{m}(z_{in}\simeq 1100)= 10^{-5}$. 
\end{itemize}
The numerical results are very sensitive to the values of $n$ and we choose random values of $n$. For the case of $n = 1$, the numerical results of GR are recovered as in Figs. \ref{fig:GRV} and \ref{fig:GRd}.  As we mentioned  in the earlier  and as presented in \cite{wei2012noether} this model has an observational valid only for the range of  $n>1.5$ for the background universe,  However, we have tried to study the behaviour of the velocity and the  density contrast for different  $n$ ranges, i.e., for $n<1 $, $n$ closes to GR and $n>1.5$. For instance, the numerical results for set I of the initial conditions,  we have noticed that the velocity and  density contrast are decaying  for values of $0.5 < n\leq 0.99 $ and  $1<n  \leq 2$, and the decaying  is faster with increasing the values of $n$.  For values of $ n>2$  we notice the growth of the velocity and the density contrast which is realistic compared to the observational expectations. The behaviour of the growth of density contrast is summarised in Table. \ref{Table1}. For set II of the initial conditions, we observe the  growth of the velocity and density contrast for values of $n>1.5$, while we have to note that the amplitudes of the velocity and density contrast are decreasing with increasing the values of $n$,  we also have to point out that we have noticed orders-of-magnitude deviations from limiting general relativistic results. We present the behaviour of the density fluctuations in Table. \ref{Table2}
\begin{table}[h!]
\caption{The behavior of $\delta(z)$ for set I.}
\begin{tabular}{lllll}
Range of $n$ & Density contarst & Expected & Remark  &  \\
 \hline
 $0.5 < n \leq 0.99 $ & decreasing &  decreasing & realistic &  \\
 \hline
 $1\leq n\geq 2 $ &  increasing & increasing & realistic &  \\
  \hline
\end{tabular}
\label{Table1}
\end{table}
\begin{table}[h!]
\caption{The behaviour of $\delta(z)$ for set II.}
\begin{tabular}{lllll}
Range of $n$ & Density contrast & Expected & Remark  &  \\
 \hline
 $0.5 < n \leq 0.99$ & decreasing &  decreasing & realistic &  \\
 \hline
 $n \geq 1 $ & increasing & increasing & realistic &  \\
 \hline
\end{tabular}
\label{Table2}
\end{table}
We evaluated the numerical results of the whole system by using the initial conditions sets I and II.   Based on these results we can conclude that the growth of the velocity and density contrast is very sensitive to the values of $n$ and the initial conditions.  The results obtained  by using conditions I are in complete disagreement with the theoretical and observational expectations. For this model we think  that  sets II of the initial conditions  give the best results as we can notice the growth of the velocity and the density contrast for values of $n>1.5$. However, the results obtained by using set II are highly nonlinear compared to the GR results. 

In the following  we present a special case for the $f(T)$ power law model. By using the definitions of the scale factor and  the Hubble parameter  Eq. \eqref{scale}, and as we mentioned earlier in the background cosmology,  we have $T= -6H^2$. Therefore,  $\dot{T}= 18H^3$,   i.e., from  the Friedmann Eq. \eqref{FR}, the term   $ \left (\dfrac{3w}{2}\tilde{\Omega}_{m}-\dfrac{3}{2}\mathcal{X}+3\mathcal{Y}\right) =0$. While $\psi =-\dfrac{3}{2}$  and  $\mathcal{Y}=- 2(n-1).$ By choosing  set I of the initial conditions,  We notice the growth of the density and velocity fluctuations increases with $n$  but they are highly nonlinear compared to the GR results  and  the GR results are recovered as shown in  Figs. \ref{fig:GRV} and \ref{fig:GRd}.
\subsection*{Case II: Quasi-static approximations}
\begin{figure}[h!]
 \begin{minipage}{0.5\textwidth}
\includegraphics[width=0.9\textwidth]{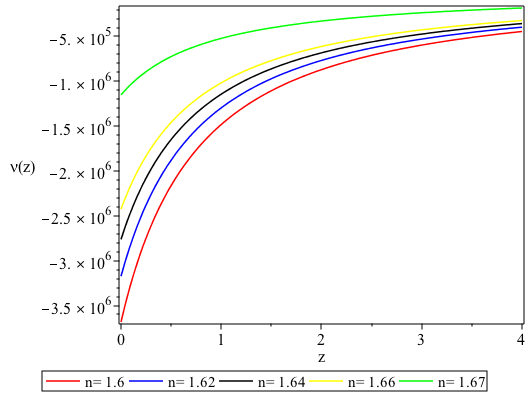}
    \caption{The growth of velocity contrast versus cosmological redshift Eq. \eqref{vs} for the power law $f(T)$ gravity model for quasi-static approximations for $n>1.5$.}
    \label{qn2}
\end{minipage} 
\qquad
 \begin{minipage}{0.45\textwidth}
\includegraphics[width=0.9\textwidth]{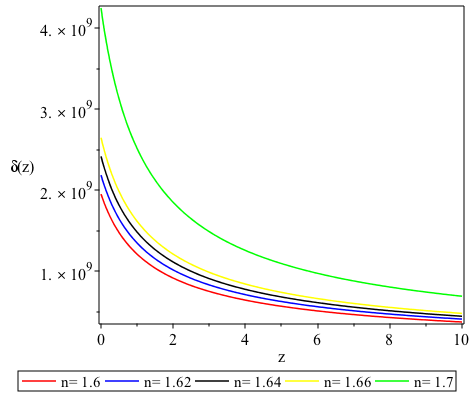}
   \caption{The growth of density contrast versus cosmological redshift Eq. \eqref{quasi188} for the power law $f(T)$ gravity model for quasi-static approximations for $n> 1.5$.}
    \label{qn1}
\end{minipage}
\end{figure}

In this sub-subsection, we present the results of the velocity and density  contrast in the quasi-static approximations.  We notice that Eq. \eqref{quasi0} is  an opened system and  Eq. \eqref{quasi2} is a closed system. We apply the same technique as GR limits and we find first the exact solution of velocity contrast and then the density contrast. So, the exact solution of Eq. \eqref{quasi2} is given as
\begin{eqnarray}\label{vs}
 &&V(z)= {\it c_5}\, \left( 1+z \right) ^{\alpha_{+}}+{\it c_6}\, \left( 1+z \right) 
^{{\alpha_{-}}}\;, \label{quasi6}
\end{eqnarray}
where \begin{eqnarray*} 
&& \alpha_{\pm} = \frac {1}{4\,\zeta} \bigg( -144\,\eta_{1}\,{\psi}^{2}+24\,\eta_{1}\,\psi+\,
\zeta_{1}\pm \\&& \sqrt {20736\,{\eta}^{2}_{1}{\psi}^{4}-6912\,{\eta}^{2}_{1}{\psi}^{3}+
288\,\zeta_{1}\,\bar{\Omega}_{m}\,\eta_{1}\,\psi+576\,{\eta}^{2}_{1}{\psi}^{2}-288\,\eta_{1}\,{
\psi}^{2}\zeta+24\,\bar{\Omega}_{m}\,{\zeta}^{2}_{1}+48\,\eta_{1}\,\psi\,\zeta_{1}+\,{
\zeta}^{2}_{1}} \bigg) \;, 
 \end{eqnarray*}
where $\eta_{1}= H\eta$ and $ \zeta _{1}= \frac{1}{H^2} \zeta$. 
After we computed the integration constants $c_5$ and $c_6$ by imposing the initial conditions, the exact solution for density contrast Eq. \eqref{quasi0} is given as
\begin{eqnarray}\label{quasi188}
&&\Delta_m(z) = {c_7}\left( 1+z \right) ^{{\frac {3\mathcal{Y}}{4}}+{\frac{3}{
4}}+{\frac {1}{4}\sqrt {9\mathcal{Y}^{2}+24\,\Omega+18\mathcal{Y}+9}}}+ {c_8} 
 \left( 1+z \right) ^{{\frac {3\mathcal{Y}}{4}}+{\frac{3}{4}}-{\frac {1}{4}
\sqrt {9\mathcal{Y}^{2}+24\,\Omega+18\mathcal{Y}+9}}}  \\ \nonumber && + \frac{ 96(1+z)^{3/2}}{(-4\alpha_{-} +6\alpha_{-}\mathcal{Y} -6\alpha_{-}+6 \bar{\Omega}_{m}+9\mathcal{Y})(-4\alpha_{+} +6\alpha_{+}\mathcal{Y} -6\alpha_{+}+6 \bar{\Omega}_{m}+9\mathcal{Y})}\times  \\ \nonumber && \Bigg(\big(\bar{\Omega}_{m} + \big(\alpha_{-}+ \frac{3}{2}\big) \big(\frac{-2\alpha_{-}}{3} +\mathcal{Y}\big) \big)\big(\big(\big( \frac{-\alpha_{+}}{2}-\frac{3}{4}\big) \zeta_{1}+ 3\beta \psi \big) \bar{\Omega}_{m} \nonumber\\&& + \alpha_{+}(( \mathcal{X}- \frac{1}{2}\big) \zeta_{1} -12\psi \big(\psi -\frac{1}{6}\big) \beta\big)\big) c_{5} (1+z)^{\alpha_{+}}  \big) +\\ \nonumber &&  \big(\bar{\Omega}_{m} +\big(\frac{3}{2}\big)+ \alpha_{+}\big) \mathcal{Y}- \frac{2\alpha^{2}_{+}}{3} -\alpha_{+}\big) \big(\big(\big( \frac{-\alpha_{+}}{2}-\frac{3}{4}\big)\zeta_{1} +3\beta \psi \big)\bar{\Omega}_{m}  \\ \nonumber &&+ \alpha_{-}\big(\big(\mathcal{X}-\frac{1}{2})\zeta_{1}- 12 \psi(\psi-\frac{1}{6}\big) \beta\big)\big)c_{6} (1+z)^{\alpha_{-}}\big)
\Bigg)\;.
\end{eqnarray}
In the following figures we present the results of the velocity and the density contrast in the quasi-static approximations Eqs . \eqref{vs} and \eqref{quasi188} for different $n$ ranges. We choose set II of the initial as it showed the most expected results for the growth of the density fluctuations. From Fig. \ref{qn1},  we clearly notice the fluctuations of the matter density highly depend on $n$, in Fig. \ref{qn1}.
\subsection*{The generalised $f(T)$ model}
In this sub-subsection, we consider the more generalised $f(T)$ model, which is constraint from solar system to be valid only for values of $\ll 1$ and it is given as \cite{wei2012noether}
\begin{eqnarray}
f(T)= T+ \mu T_{0}\left(\frac{T}{T_{0}}\right)^{n}\;.
\end{eqnarray}
From Eq. \eqref{rho111}, the parameter $ \mu$ reads as
\begin{eqnarray}
\mu= \dfrac{1-\Omega_{m}}{(1-2n)}, \hspace*{ .4cm} n \neq 0.5 \;.
\end{eqnarray}
The background quantities $\mathcal{X}$, $\tilde{\Omega}_{m}$ and $\mathcal{Y}$ as defined in Eq. \eqref{defi1} become 
\begin{eqnarray}
&&\tilde{\Omega}_{m}=  (1-\mathcal{X})\;,\\&&
\mathcal{Y}= \Big\lbrace\dfrac{12H^{2} n(n-1)(1+\mathcal{X})\mu (T/T_0)^{n-2}}{T_{0}(1+\mu n(T/T_0)^{n-1})}\Big\rbrace \times \nonumber \\&& \Big\lbrace\dfrac{1}{\Big(1+\dfrac{ 24 H^{2}n(n-1) \mu (T/T_{0})^(n-2)}{T_{0}(1+n\mu(T/T_{0})^{n-1})}\Big)}\Big\rbrace\;.
\end{eqnarray}
\subsection*{Case I: Solving the whole system for the generalised $f(T)$ model}
\begin{figure}[h!]
 \begin{minipage}{0.48\textwidth}
\includegraphics[width=0.9\textwidth]{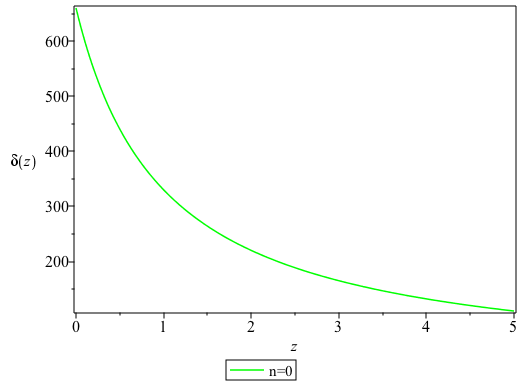}
    \caption{The growth of velocity contrast versus cosmological redshift for the generalized $f(T)$ model for the system of Eqs. \eqref{00000} - \eqref{ddot10100000} for $n=0 $.}
    \label{Figuref5}
\end{minipage} 
\qquad
\begin{minipage}{0.5\textwidth}
\includegraphics[width=0.9\textwidth]{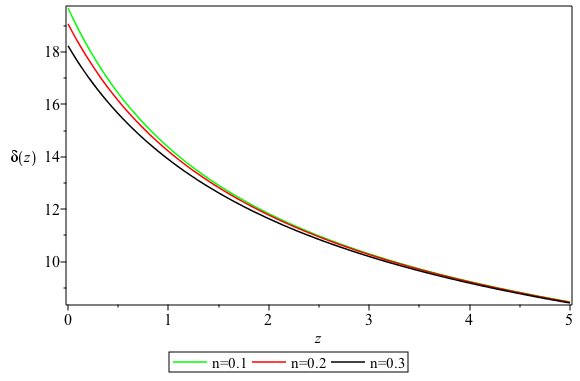}
    \caption{The growth of the density contrast versus cosmological redshift for the  generalized $f(T)$ model for the system of Eqs. \eqref{00000} - \eqref{ddot10100000} for $n>0$, $\tilde{\Omega}_{m}= 0.32$ and $k= 10^5$.}
     \label{Figuref7}
\end{minipage}
\end{figure}

Based on the  above definitions, the numerical results of this system of the first-order equations \eqref{00000} - \eqref{ddot10100000}  are presented in such way that we set our free parameter $\tilde{\Omega}_{m}= 0.32$ based on the observational expectations.  We set the initial conditions at $ V _{in}= V(z_{in}\simeq 1100)= 0$,  $ \mathcal{F} _{in}= \mathcal{F}(z_{in}\simeq 1100)= 0$ , $\mathcal{B} _{in}= \mathcal{B}(z_{in}\simeq 1100)= 0$, $\mathcal{A} _{in}= \mathcal{A}(z_{in}\simeq 1100)= 0$,  $Z _{in}= Z(z_{in}\simeq 1100)= 0$  and $\Delta _{in}= \Delta _{m}(z_{in}\simeq 1100)= 10^{-5}$.  As we mentioned earlier only at this specific choice of the initial conditions, the $k$ dependence does not make any difference in the behaviour or the amplitude  of the density contrast. The  numerical results of this system of Eqs. \eqref{00000} - \eqref{ddot10100000}  are presented in Figs. \ref{Figuref5} - \ref{Figuref7}, For the case of $n = 0$, the numerical results of GR are recovered.   We have noticed the growth of the density contrast  for values of $n$ close to $0$  for different values of $k$ and  as we mentioned  before   the $k$ dependence does not make any difference in the behaviour or the amplitude  of the density contrast only at this specific choice of the initial conditions.
\newline
 In the following we will solve the whole system of the second-order perturbation equations,  evaluating the numerical solutions from Eqs. \eqref{non-quasi111} - \eqref{non-quasi3333}  simultaneously to analyse the density and velocity fluctuations as a function of redshift.  For further analysis, we evaluated  the numerical results for different initial conditions as we did for the power law model. The numerical results of this system of  Eqs. \eqref{non-quasi111} - \ref{non-quasi3333}   for set I of the initial conditions , the GR results are recovered  as in Figs. \ref{fig:GRV} and \ref{fig:GRd} for $n=0$ and  we noticed the growth of the matter density and velocity for very small values of $n$ but the results are  still highly nonlinear compared to the GR results.
 \newline
As we mentioned this model is tested to be valid for very small values of $n\ll1$, we tried set II  of the initial conditions and we noticed the decay of the velocity and the density fluctuations for values of $n\ll1$. We can conclude  based on the results obtained that this model is viable for our study only if we choose  the initial conditions for the first derivatives of the  variables $\Delta_{m} $, $V$ and $\mathcal{F}$  to be  zero at the initial red-shift. In other words, set II  of the initial conditions does not provide good results for this model. We also found that the exact solutions in the quasi-static approximations for this model could not be evaluated.
\section{Conclusions}\label{sec8}
This work presented a detailed analysis of scalar cosmological perturbations in the $f(T)$ gravity theory using the $1 + 3$ covariant gauge-invariant approach. We explored the integrability conditions of the so-called quasi-Newtonian cosmological models in the context of $f(T)$ gravity, a first such study to the best of our knowledge. We showed that for such cosmological models to exist, they must satisfy certain integrability conditions on the generalised Einstein field equations. The two integrability conditions derived and presented here allow us to describe a consistent evolution of the linearised field equations of quasi-Newtonian universes. We defined the gauge-invariant variables and derived the corresponding evolution equations. We derived the complete set of the first- and the second-order evolution equations of these perturbations 
These results agree with the GR results when  $f(T)= T$.  We studied the behaviour of matter energy density perturbations with redshift for different ranges of the model-defining parameter $n$ by considering two of the $f(T)$ gravity models. We employed the so-called quasi-static approximation technique on small scales. Some of the specific highlights of this work are as follows: in the first model $f(T)= \mu T_{0}(\frac{T}{T_{0}})^n\;,$ we presented the ranges of values of $n$ for which the perturbation amplitudes $\delta(z)$  grow or decay. For instance, the numerical solution  of the first-order perturbation equations  shows the growth of the density contrast for values of $n>1.5$ only at one particular choice of initial conditions  
and that the wave number $k$ has no effect on the behaviour of the density contrast (see  Fig. \ref{Figuref3}). We also found numerical solution of the whole system of second-order equations and for different sets of initial conditions. For instance,  for set I we observed unrealistic behaviour of the density  perturbations for $0.5\leq n \leq 0.99$ and for $ 1<n \leq 2$,  but for values of $ n>2$, we found that the growth of the density contrast was consistent with observational expectations. The results obtained in the first model generalised existing GR results as it is shown in  Figs. \ref{fig:GRV} and \ref{fig:GRd}. We also noticed that the behaviour of the density fluctuations is highly dependent on our choice of the initial conditions. For instance,  for set II of the initial conditions, we found results which are highly non-linear compared to expected GR results. The results obtained  using the first set of initial conditions (Set I)  are in complete disagreement with the theoretical and observational expectations.  On the other hand, applying the quasi-static approximation to his model, the growth of the density perturbations can be explored for the range of $n>1.5$  for Set I of the initial conditions. By comparing the complete system with the quasi-static approximation, for the same values of $n$ and the same  choice of the initial conditions, we found that there is  a big difference in the amplitudes of the density fluctuations. In conclusion, the quasi-static approximation appears to be not applicable for this model.

In the second case: $f(T)= T+ \mu T_{0} (T/T_{0})^{n}$, with the value of $n$ constrained by solar system tests to be in the range $n \ll 1$, we obtained a growth in the velocity perturbations and the density contrast $\delta (z)$  for  $0 \leq n \leq 0.4$. We considered a range of possible initial conditions, and found that this model is only viable when the first derivatives of the variables $\Delta_{m}$, $V$ and $\mathcal{F}$ are equal to zero.  
In conclusion, we have developed a covariant framework for studying quasi-Newtonian cosmologies in the $f(T)$ gravity theory. Future studies in this direction using more realistic (i.e., in terms of cosmological and astrophysical viability) models of the theory using actual astronomical data might shed more light on whether such a framework can be a viable cosmological alternative to late-time cosmology.
 \section*{Acknowledgments}
HS gratefully acknowledges the financial support from the Mwalimu Nyerere African Union scholarship and the National Research Foundation (NRF) free-standing scholarship with a grant number  112544.  SS gratefully acknowledges financial support from Entoto Observatory and Research Center and
Ethiopian Space Science and Technology Institute, Wolkite University as well as the hospitality of the Physics Department of North-West University (NWU) during the conceptualization stage of this project. AA acknowledges that this work is based
on the research supported in part by the NRF of South Africa with grant number 112131. 
\newpage
\appendix
\section*{Appendix}
\renewcommand{\theequation}{\thesection.\arabic{equation}}
\renewcommand{\thesection}{A} 
Some of the following linearised identities which hold for all scalars $f$, vectors $V_a$ and tensors $S_{ab}=S_{\langle ab\rangle}$, have been used in this paper:
\begin{eqnarray}\label{a1}
&&\eta^{abc}\tilde{\nabla}_{b}\tilde{\nabla}_{c}f=0\; ,\\
&&\label{a2}
\left( \tilde{\nabla}_{\langle a}\tilde{\nabla}_{b\rangle}f\right)^{\cdot}=\tilde{\nabla}_{\langle a}\tilde{\nabla}_{b\rangle} \dot{f} -\dfrac{2}{3} \theta \tilde{\nabla}_{\langle a}\tilde{\nabla}_{b\rangle}f +\dot{f}\tilde{\nabla}_{\langle a}A_{b\rangle}\; ,\\
&&\label{a3}
\left( \tilde{\nabla}_{a}f\right)^{\cdot}= \tilde{\nabla}_{a}\dot{f}-\dfrac{1}{3}\theta\tilde{\nabla}_{a}f+ \dot{f}A_{a}\; ,\\
&&\label{a4}
\tilde{\nabla}^{b}\tilde{\nabla}_{<a}A_{b>}= \dfrac{1}{2}\tilde{\nabla}^{2}A_{a}+\dfrac{1}{6}\tilde{\nabla}_{a}\tilde{\nabla}^{c}A_{c}+ \dfrac{1}{3}(\rho- \dfrac{1}{3} {\theta}^{2})A_{a}\;,\\
&&\label{a5}
\left(\tilde{\nabla}^{2}f\right)^{\cdot}=\tilde{\nabla}^{2}\dot{f}-\dfrac{2}{3}\theta \tilde{\nabla}^{2}f+ \dot{f}\tilde{\nabla}^{a}A_{a}\; ,\\
&&\label{a6}
\tilde{\nabla}_{[a}\tilde{\nabla}_{b]}v_{c}=\dfrac{1}{3}\left(\dfrac{1}{3}\theta^{2}-\rho\right)v_{[a}h_{b]c}\; ,\\
&&\label{a7}
\tilde{\nabla}_{[a}\tilde{\nabla}_{b]}S^{cd}=\dfrac{2}{3}\left(\dfrac{1}{3}\theta^{2}-\rho\right)S_{[a}^{(c}h_{b]}^{d)}\; ,\\
&&\label{a8}
\tilde{\nabla}^{a}\left( \eta_{abc}\tilde{\nabla}^{b}v^{c}\right)=0\; ,\\
&&\label{a9}
\tilde{\nabla}_{b}\left( \eta^{cd\langle a}\tilde{\nabla}_{c}S^{b\rangle}_{d}\right)= \dfrac{1}{2}\eta^{abc}\tilde{\nabla}_{b}\left(\tilde{\nabla}_{d}S^{d}_{c}\right)\; ,\\
&&\label{a10}
\tilde{\nabla}^{2}(\tilde{\nabla}_{a}f)= \tilde{\nabla}_{a}(\tilde{\nabla}^{2}f)+\dfrac{2}{3}(\rho-\dfrac{1}{3}\theta^{2})\tilde{\nabla}_{a}f+ 2\dot{f}\eta_{abc}\tilde{\nabla}^{b} \omega^{c}\;.
\end{eqnarray}

\section*{References}

\bibliography{references}

\providecommand{\newblock}{}
\begin{thebibliography}{10}
\expandafter\ifx\csname url\endcsname\relax
  \def\url#1{{\tt #1}}\fi
\expandafter\ifx\csname urlprefix\endcsname\relax\def\urlprefix{URL }\fi
\providecommand{\eprint}[2][]{\url{#2}}

\bibitem{riess1998observational}
Riess A~G {\em et~al.\/} 1998 {\em The Astronomical Journal\/} {\bf 116} 1009

\bibitem{perlmutter1998discovery}
Perlmutter S {\em et~al.\/} 1998 {\em Nature\/} {\bf 391} 51--54

\bibitem{copeland2006dynamics}
Copeland E~J, Sami M and Tsujikawa S 2006 {\em International Journal of Modern
  Physics D\/} {\bf 15} 1753--1935

\bibitem{nojiri2007introduction}
Nojiri S and Odintsov S~D 2007 {\em International Journal of Geometric Methods
  in Modern Physics\/} {\bf 4} 115--145

\bibitem{liu2012energy}
Liu D and Reboucas M 2012 {\em Physical Review D\/} {\bf 86} 083515

\bibitem{paliathanasis2018stability}
Paliathanasis A, Said J~L and Barrow J~D 2018 {\em Physical Review D\/} {\bf
  97} 044008

\bibitem{capozziello2001geometric}
Capozziello S, Lambiase G and Stornaioloi C 2001 {\em Annalen der Physik\/}
  {\bf 10} 713--727

\bibitem{sahlu2019accelerating}
Sahlu S {\em et~al.\/} 2019 {\em Proceedings of the International Astronomical
  Union\/} {\bf 15} 397--399

\bibitem{sahlu2019chaply}
Sahlu S {\em et~al.\/} 2019 {\em The European Physical Journal C\/} {\bf 79}
  749

\bibitem{sahlu2020scalar}
Sahlu S {\em et~al.\/} 2020 {\em European Physical Journal C\/} {\bf 80} 1--19

\bibitem{tsamparlis1979cosmological}
Tsamparlis M 1979 {\em Physics Letters A\/} {\bf 75} 27--28

\bibitem{einstein1925unified}
Einstein A 1925 {\em Session Report of Prussian Acad. Sci\/}  414--419

\bibitem{einstein1928riemannian}
Einstein A 1928 {\em Sitz. Preuss. Akad. Wiss\/} {\bf 217}

\bibitem{einstein1928new}
Einstein A 1928 {\em Sitzungsberichte der Preussischen Akademie der
  Wissenschaften. Physikalisch-Mathematische Klasse\/}  223--227

\bibitem{unzicker2005translation}
Unzicker A and Case T 2005 {\em arXiv preprint physics/0503046\/}

\bibitem{turnbull1924invarianten}
Turnbull H 1924 {\em The Mathematical Gazette\/} {\bf 12} 122--124

\bibitem{3}
Arcos H and Pereira J 2004 {\em International Journal of Modern Physics D\/}
  {\bf 13} 2193--2240

\bibitem{sciama1962analogy}
Sciama D~W 1962 {\em Recent developments in general relativity\/}  415

\bibitem{hehl1995metric}
Hehl F~W, McCrea J~D, Mielke E~W and Ne'eman Y 1995 {\em Physics Reports\/}
  {\bf 258} 1--171

\bibitem{carvalho2004torsion}
Carvalho~de Andrade V 2004 Torsion as alternative to curvature in the
  description of gravitation {\em Mathematical Methods in Physics\/} p~28

\bibitem{aldrovandi2012teleparallel}
Aldrovandi R and Pereira J~G 2012 {\em Teleparallel gravity: an introduction\/}
  vol 173 (Springer Science \& Business Media)

\bibitem{darabi2015geodesic}
Darabi F, Mousavi M and Atazadeh K 2015 {\em Physical Review D\/} {\bf 91}
  084023

\bibitem{arcos2004torsion}
Arcos H, De~Andrade V and Pereira J 2004 {\em International Journal of Modern
  Physics D\/} {\bf 13} 807--818

\bibitem{krvsvsak2017variational}
Kr{\v{s}}{\v{s}}{\'a}k M 2017 {\em arXiv preprint arXiv:1705.01072\/}

\bibitem{linder2010einstein}
Linder E~V 2010 {\em Physical Review D\/} {\bf 81} 127301

\bibitem{iorio2012solar}
Iorio L and Saridakis E~N 2012 {\em Monthly Notices of the Royal Astronomical
  Society\/} {\bf 427} 1555--1561

\bibitem{li2011f}
Li B, Sotiriou T~P and Barrow J~D 2011 {\em Physical Review D\/} {\bf 83}
  064035

\bibitem{li2011large}
Li B, Sotiriou T~P and Barrow J~D 2011 {\em Physical Review D\/} {\bf 83}
  104017

\bibitem{lifshitz1946gravitational}
Lifshitz E~M 1946 {\em Zhurnal Eksperimentalnoi i Teoreticheskoi Fiziki\/} {\bf
  16} 587--602

\bibitem{bardeen1980gauge}
Bardeen J~M 1980 {\em Physical Review D\/} {\bf 22} 1882

\bibitem{kodama1984cosmological}
Kodama H and Sasaki M 1984 {\em Progress of Theoretical Physics Supplement\/}
  {\bf 78} 1--166

\bibitem{bertschinger2000cosmological}
Bertschinger E 2000 {\em arXiv preprint astro-ph/0101009\/}

\bibitem{bruni1992cosmological}
Bruni M, Dunsby P~K and Ellis G~F 1992 {\em The Astrophysical Journal\/} {\bf
  395} 34--53

\bibitem{dunsby1991gauge}
Dunsby P~K 1991 {\em Classical and Quantum Gravity\/} {\bf 8} 1785

\bibitem{dunsby1992covariant}
Dunsby P~K, Bruni M and Ellis G~F 1992 {\em The Astrophysical Journal\/} {\bf
  395} 54--74

\bibitem{gidelew2013beyond}
Gidelew A~A 2013 {\em Beyond the concordance cosmology\/} Ph.D. thesis
  University of Cape Town

\bibitem{hwang1991large}
Hwang J~C 1991 {\em The Astrophysical Journal\/} {\bf 380} 307--314

\bibitem{mukhanov1992theory}
Mukhanov V~F, Feldman H~A and Brandenberger R~H 1992 {\em Physics Reports\/}
  {\bf 215} 203--333

\bibitem{hawking1966perturbations}
Hawking S 1966 {\em The Astrophysical Journal\/} {\bf 145} 544

\bibitem{ellis1989covariant}
Ellis G~F and Bruni M 1989 {\em Physical Review D\/} {\bf 40} 1804

\bibitem{challinor2000microwave}
Challinor A 2000 {\em Classical and Quantum Gravity\/} {\bf 17} 871

\bibitem{abebe2012covariant}
Abebe A, Abdelwahab M, De~la Cruz-Dombriz {\'A} and Dunsby P~K 2012 {\em
  Classical and Quantum Gravity\/} {\bf 29} 135011

\bibitem{cai2016f}
Cai Y~F, Capozziello S, De~Laurentis M and Saridakis E~N 2016 {\em Reports on
  Progress in Physics\/} {\bf 79} 106901

\bibitem{carloni2010conformal}
Carloni S, Elizalde E and Odintsov S 2010 {\em General Relativity and
  Gravitation\/} {\bf 42} 1667--1705

\bibitem{ellis1999cosmological}
Ellis G~F and Van~Elst H 1999 Cosmological models {\em Theoretical and
  Observational Cosmology\/} (Springer) pp 1--116

\bibitem{castaneda2016some}
Castaneda C {\em et~al.\/} 2016 {\em Some Aspects in Cosmological Perturbation
  Theory and {${f(R)}$} Gravity\/} Ph.D. thesis Dissertation, Bonn, Rheinische
  Friedrich-Wilhelms-Universit{\"a}t Bonn, 2016

\bibitem{pasmatsiou2017kinematics}
Pasmatsiou K, Tsagas C~G and Barrow J~D 2017 {\em Physical Review D\/} {\bf 95}
  104007

\bibitem{ehlers2007ak}
Ehlers J 2007 {\em Pramana\/} {\bf 69} 7--14

\bibitem{carloni2008evolution}
Carloni S, Dunsby P and Troisi A 2008 {\em Physical Review D\/} {\bf 77} 024024

\bibitem{ehlers1993contributions}
Ehlers J 1993 {\em General Relativity and Gravitation\/} {\bf 25} 1225--1266

\bibitem{maartens98}
Maartens R 1998 {\em Physical Review D\/} {\bf 58} 124006

\bibitem{stewart1974perturbations}
Stewart J~M and Walker M 1974 {\em Proceedings of the Royal Society of London.
  A. Mathematical and Physical Sciences\/} {\bf 341} 49--74

\bibitem{abebe2016integrability}
Abebe A, Dunsby P~K~S and Solomons D 2016 {\em International Journal of Modern
  Physics D\/} {\bf 26} 1750054

\bibitem{maartens1997density}
Maartens R and Triginer J 1997 {\em Physical Review D\/} {\bf 56} 4640

\bibitem{abebe2015irrotational}
Abebe A and Elmardi M 2015 {\em International Journal of Geometric Methods in
  Modern Physics\/} {\bf 12} 1550118

\bibitem{maartens1998newtonian}
Maartens R, Lesame W~M and Ellis G~F 1998 {\em Classical and Quantum Gravity\/}
  {\bf 15} 1005

\bibitem{van1998quasi}
Van~Elst H and Ellis G~F 1998 {\em Classical and Quantum Gravity\/} {\bf 15}
  3545

\bibitem{maartens1997linearization}
Maartens R 1997 {\em Physical Review D\/} {\bf 55} 463

\bibitem{Sami:2018lxt}
Abdulrahman H~S and Abebe A 2018 {\em LHEP\/} {\bf 1} 21--27 (\textit{Preprint}
  \eprint{1802.06778})

\bibitem{van97}
van Elst H, Uggla C, Lesame W~M, Ellis G and Maartens R 1997 {\em Classical and
  Quantum Gravity\/} {\bf 14} 1151

\bibitem{ellis1990density}
Ellis G, Bruni M and Hwang J 1990 {\em Physical Review D\/} {\bf 42} 1035

\bibitem{carloni2006gauge}
Carloni S, Dunsby P~K and Rubano C 2006 {\em Physical Review D\/} {\bf 74}
  123513

\bibitem{maartens1998covariant}
Maartens R 1998 {\em Physical Review D\/} {\bf 58} 124006

\bibitem{smoot1992structure}
Smoot G~F, Bennett C~L, Kogut A, Wright E, Aymon J, Boggess N, Cheng E,
  De~Amici G, Gulkis S, Hauser M {\em et~al.\/} 1992 {\em The Astrophysical
  Journal\/} {\bf 396} L1--L5

\bibitem{wei2012noether}
Wei H, Guo X~J and Wang L~F 2012 {\em Physics Letters B\/} {\bf 707} 298--304

\bibitem{abebe2013large}
Abebe A, de~la Cruz-Dombriz {\'A} and Dunsby P~K 2013 {\em Physical Review D\/}
  {\bf 88} 044050

\end{thebibliography}
\bibliographystyle{iopart-num}

\end{document}